\documentclass[fleqn,10pt]{wlscirep}
\usepackage[utf8]{inputenc}
\usepackage[T1]{fontenc}
\usepackage{blkarray}
\usepackage{graphbox}
\title{Evolutionary games on multilayer networks: coordination and equilibrium selection}

\author[1,2,*]{Tomasz Raducha}
\author[2]{Maxi San Miguel}
\affil[1]{Grupo Interdisciplinar de Sistemas Complejos (GISC), Departamento de Matem\'aticas, Universidad Carlos III de Madrid, Legan\'es, Spain}
\affil[2]{Instituto de F\'isica Interdisciplinar y Sistemas Complejos IFISC (CSIC-UIB), Palma, Spain}

\affil[*]{tjan@math.uc3m.es}

\keywords{coordination games, multilayer networks, multiplex networks, game theory, evolutionary games, equilibrium selection}

\begin{abstract} 
We study mechanisms of synchronisation, coordination, and equilibrium selection
in two-player coordination games on multilayer networks.
We apply the approach from evolutionary game theory with three
possible update rules: the replicator dynamics (RD),
the best response (BR), and the unconditional imitation (UI).
Players interact on a two-layer random regular network.
The population on each layer plays a different game,
with layer I preferring the opposite strategy to layer II.
We measure the difference between the two games played on the layers
by a difference in payoffs $\Delta S$ while the inter-connectedness is measured 
by a node overlap parameter $q$.
We discover a critical value $q_c(\Delta S)$ below which
layers do not synchronise. For $q>q_c$ in general both layers
coordinate on the same strategy. Surprisingly, there is a symmetry breaking
in the selection of equilibrium -- for RD and UI there is
a phase where only the payoff-dominant equilibrium is selected.
Our work is an example of previously observed differences between the update rules on a single network.
However, we took a novel approach with the game being played
on two inter-connected layers.
As we show, the multilayer structure enhances the abundance of the
Pareto-optimal equilibrium in coordination games with imitative update rules.
\end{abstract}
\begin{document}

\flushbottom
\maketitle

\thispagestyle{empty}


\section*{Introduction}

Spontaneous emergence of coordination between people or animals, without
external control, is a remarkable phenomenon that can be crucial for
optimal functioning or even survival of the population
\cite{king2009origins,conradt2009group,courchamp2002small}.
In some circumstances individuals face making a choice between two
or more possible actions, called strategies. It often happens that the best outcome
for everyone can be obtained only if we choose the same strategy
as our neighbours. In game theory such situation is referred to
as coordination game
\cite{weidenholzer2010coordination,antonioni2013coordination,mazzoli2017equilibria,antonioni2014global}.
Additionally, it might matter under which strategy
the population coordinates.
One action can lead to higher prosperity than the other, what
is modelled by different strategies having different payoffs.
Conditions required to coordinate have been scrutinised under various
assumptions and for numerous environments, yet there are still unanswered
questions. Here, we study coordination and equilibrium selection in
games on multilayer networks.

People interact in multiple contexts and through different media.
One  natural way to represent it in a strict manner is by using a
multilayer network \cite{boccaletti2014structure,kivela2014multilayer,battiston2014structural,battiston2017new,aleta2019multilayer}.
Each layer is a separate network of interactions
in a given context. For example, we interact with each other in work place,
at home, online etc. In principle, the pattern of interactions can be
different in every layer resulting in a different network topology.
Additionally, some layers can be hidden \cite{gajewski2021detecting}.
In multilayer networks, if a node exists
in many layers, it represents the same person, which often acts
similarly in every context. It is therefore
connected between layers to itself
via inter-layer links, which provide coupling between the layers.
It is important to note that, if a system has a multilayer structure,
it can not be simply reduced to a single-layer graph without
changing the dynamics \cite{diakonova2016irreducibility}.
Hence, the scrutiny of layered systems is highly relevant.

We use the approach from evolutionary
game theory \cite{sigmund1999evolutionary,axelrod1981evolution,nowak1992evolutionary,nowak2006five}
to analyse synchronisation 
between the layers and equilibrium selection in coordination games.
Coordination games have been studied in depth on single layer networks,
a comprehensive literature review can be found here
\cite{raducha2022coordination}. From previous results it is
worth mentioning the KMR model which
explored the equilibrium selection in populations equivalent to
complete graphs with the best response update rule \cite{kandori1993learning}.
The risk-dominant equilibrium was always evolutionary
favoured in the model and several extensions did not find any deviation from
this behaviour \cite{young1993evolution,youngindividual,ellison2000basins,peski2010generalized}. That outcome is preserved also on a circular network
\cite{ellison1993learning}, unless the unconditional imitation
is used to update strategies \cite{alos2006imitation}. In general,
imitative update rules can favour Pareto-efficiency over risk dominance
\cite{ohtsuki2006replicator,roca2009evolutionary}.
However, it can only happen in sparse networks -- in a complete graph 
risk-dominant equilibrium is always selected\cite{raducha2022coordination}.

Evolutionary games were also extended to multilayer networks
\cite{wang2015evolutionary}.
Prisoner's dilemma was studied on many layers with a possibility
of using different strategies on different layers.
The strategy was updated according to replicators dynamics,
but using the collective payoff from all
layers \cite{gomez2012evolution,matamalas2015strategical}.
It was also studied together with the stag hunt, the harmony game,
and the snow drift on two-layer networks with the game being
played on one layer and strategy imitation on the other
\cite{wang2014degree}. Additionally, the same games
on one layer were mixed with opinion dynamics and
social influence on the second layer \cite{amato2017interplay}.
The idea of separating the group in which we play the game
from the one where we learn or imitate the strategy had been
already studied before within a single network
\cite{alos2014imitation,cui2016collaboration,khan2014coordination,alos2021efficient}.
The public goods game
\cite{tomassini2021computational,giardini2021gossip,maciel2021framing}
was considered on two \cite{wang2013interdependent}
and more layers \cite{battiston2017determinants} with the game
being played on each layer.
Interestingly, in some of the mentioned research the multilayer structure
was said to enhance cooperation
\cite{gomez2012evolution,amato2017interplay,wang2013interdependent}.
Finally, coordination games were also investigated on multilayer
networks. The pure coordination game on one layer was
coupled with social dynamics and coevolution on the other, leading
to a possible segregation \cite{lipari2019investigating}.
A particular version of the general coordination game
was studied on two interconnected layers, with the strategy being imitated on the layers
and the game played between the layers \cite{lugo2015learning,lugo2020local,gonzalez2019coordination}.
Similarly to single-layer networks,
the unconditional imitation and smaller degree favoured
the Pareto-optimal equilibrium.
However, the body of work on coordination games on multilayer
networks is still very limited and consists of particular cases
of more complex models mixed with opinion dynamics.
Moreover, different works consider different update rules and it is
difficult to judge to which extent results are determined by
the multilayer structure, the particular payoff matrix,
or the chosen update rule. Comparison between different
update rules is necessary.
For these reasons, we provide a broader analysis
of different payoff matrices laying within
the coordination games scope together with
three different update rules.

We focus on the two-player general coordination game\cite{raducha2022coordination} described by a $2 \times 2$ payoff matrix:
\begin{equation}
\begin{blockarray}{ccc}
 & $A$ & $B$  \\
\begin{block}{c(cc)}
  $A$ & 1 & S  \\
  $B$ & T & 0  \\
\end{block}
\end{blockarray}~,
\label{eqn:matrix_most_general}
\end{equation}
where A and B are available strategies, while $T$ and $S$ are
parameters defining payoffs. By definition, coordination games
must fulfil conditions $T<1$ and $S<0$.
A game described by such payoff matrix contains a social
dilemma. Obviously, the most rewarding outcome is obtained if both
players choose the same strategy, but there is a hidden trade off
between security and profit. Clearly, the highest possible profit is made
when both play the strategy A, hence it is called the payoff-dominant
or Pareto-optimal strategy. On the other hand, the risk-dominant strategy
is the best choice in the lack of knowledge, i.e. it is the strategy that
results in the highest average payoff assuming that the opponent
will play either way with the same probability
\cite{harsanyi1988general}.
It is easy to check that for $T<S+1$ the strategy A is risk-dominant,
and for $T>S+1$ the strategy B is be risk-dominant. This calculation
provides a theoretical line $T=S+1$ at which risk dominance changes.
When all players coordinate on one of these strategies we 
refer to such state as a payoff-dominant or risk-dominant equilibrium.

In the evolutionary game theory the game evolves because the players
update their strategies after interacting and observing their peers.
It is well known that the update rule is as important as the payoff
matrix in defining the end result of the game
\cite{raducha2022coordination,ohtsuki2006replicator,roca2009evolutionary,xia2012role,szolnoki2018dynamic,danku2018imitate,poncela2016humans}.
Multiple update
rules have been proposed in the literature 
\cite{szabo2007evolutionary,pangallo2021towards,blume1993statistical,traulsen2007pairwise}. We focus on three
well established ones: the replicator dynamics (RD)
\cite{schuster1983replicator,hammerstein1994game,nowak2004evolutionary}, the best response
(BR)
\cite{kandori1993learning,young1993evolution,blume1995statistical,ellison1993learning,sandholm1998simple,buskens2008consent}, and the unconditional imitation (UI)
\cite{nowak1992evolutionary,vilone2012social,vilone2014social,lugo2015learning,gonzalez2019coordination,lugo2020local}. It is important to note
that RD and UI are imitative in nature, as players adapt the strategy
of one of the neighbours. BR on the other hand is a strategical update rule
which requires from the player knowledge abut the payoff matrix.
Another distinction between the update rules is their determinism
-- BR and UI are deterministic, meaning that the same configuration
will always lead to the same strategy being chosen, while RD is a
probabilistic update rule. See Methods section for more details.

\begin{figure}[ht]
\centerline{
\includegraphics[scale=1,align=c]{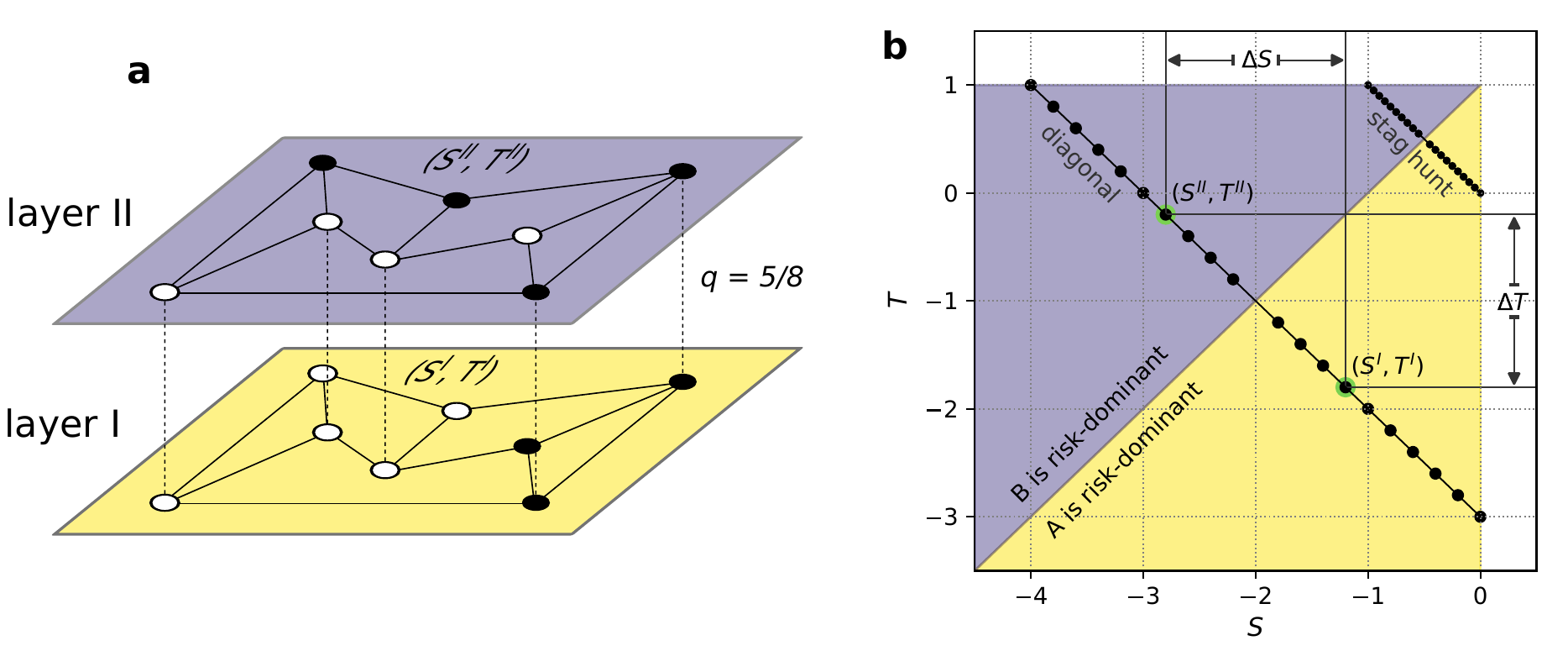}
}
\caption{
(a) Schematic representation of a miniature of multilayer network used in our
simulations. Both layers have the same topology of a random regular
graph with $N=8$ nodes of degree $k=3$ each
and a fraction $q=5/8$ of nodes
is shared between the layers. Shared nodes are connected by inter-layer
connections (dashed lines). The node overlap $q$ is the number of
shared nodes divided by $N$. White nodes play the strategy A
and black ones play the strategy B. Shared nodes always have the same
state on both layers. Each layer has a specific payoff matrix
given by $(S^I, T^I)$ and $(S^{II}, T^{II})$.
(b) Diagram of the $S$-$T$ parameter space showing parametrisation
of the layers. Each circle on the diagonal lines represents a game
played on one of the layers. Examplary values
of $(S^I, T^I)$ and $(S^{II}, T^{II})$ are highlighted in green
with $\Delta S$ and $\Delta T$ illustrated. On layer I the strategy A is always
risk-dominant (yellow area),
and on layer II the strategy B is always risk-dominant (purpule area).
Risk-dominance changes at the line $T=S+1$. }
\label{fig:multilayer_diagram}
\end{figure}

It was shown that on a single-layer network the risk aversion
is usually stronger than the drive to profit. Therefore, on complete
graphs the risk-dominant equilibrium is always obtained. For sparse
networks under unconditional imitation the system
can favour the Pareto-optimal
equilibrium over the risk-dominant one, but only for a limited
range of parameters \cite{raducha2022coordination}.
For RD and BR, however, the risk-dominant equilibrium is always selected.
In general, local effects were shown to be more important
for update rules
which have an imitative nature, such as unconditional imitation.
\cite{alos2006imitation,ohtsuki2006replicator,roca2009evolutionary}.
A natural question is which equilibrium, if any, will be chosen
when the population is placed on a multilayer network
with two layers on opposite sides of the $T=S+1$ risk-dominance transition line.
In other words,
on layer I agents play a game where the strategy A is risk-dominant
and on layer II a game where the strategy B is risk-dominant.
We investigate it by means of numerical simulations.

We study a population of players participating in two games on
a multilayer network with
two inter-connected layers, as depicted in
Figure~\ref{fig:multilayer_diagram}.
Both layers have the same number of nodes $N$.
If a node is connected
to itself between the layers via an inter-link, it plays the same
strategy in the two layers. The fraction of nodes connected (or shared) between the layers
is controlled by a parameter $q \in [0,1]$, called node overlap
or degree of multiplexity \cite{diakonova2014,diakonova2016irreducibility}. There are $Nq$ inter-layer connections. For $q=0$ the two layers
are effectively two independent networks, for $q=1$ the layers
are equivalent to one network (every node has the same state on each layer all the time)
playing each game half of the times. The edge overlap \cite{battiston2014structural} is kept constant
and equal to 1
with both layers having the same topology, since we did not observe
any change under varying edge overlap. We use random regular graphs
\cite{newmannetworks}. See Methods for more details on our simulations.

Players on each layer are engaged in different games, i.e. parameters
$S^\beta$ and $T^\beta$, $\beta \in \{  \textrm{I, II}\}$, 
defining the payoff matrix have different values
on each layer. In order to give the same relevance to both layers,
their preferences towards one of the equilibria are set to be
equally strong. This is achieved by choosing the points $(S^I, T^I)$
and $(S^{II}, T^{II})$ equally distant from the $T=S+1$ line,
as visible in Figure~\ref{fig:multilayer_diagram}. Another choice to make
is the angle between the $T=S+1$ line and the line created by
points $(S^I, T^I)$ and $(S^{II}, T^{II})$. We focus on cases
where all points lay on a line $T^\beta=-S^\beta +C$,
where $C$ is a constant
(see Supplementary Material for other cases).
This is because only then the average payoffs $\langle \Pi^I \rangle$
and $\langle \Pi^{II} \rangle$ of both layers are equal, therefore
games are truly symmetrical. We analyse the case of $T^\beta=-S^\beta -3$,
which we call \textit{diagonal}, and $T^\beta=-S^\beta$
where all games are variants of
the well known \textit{stag hunt} \cite{skyrms2001stag,skyrms2004stag}.
Note, that the stag hunt game can be obtained for different values of $C$ and that both cases
are ,,diagonal'' in the sense that they have the same slope.
Nevertheless, we call the case of $C=-3$ \textit{diagonal} and
$C=0$ \textit{stag hunt} to easily distinguish them in the discussion of results that
follows in the manuscript.
In both cases we cover with
the parameters $S$ and $T$ the whole width of the general
coordination game area (see Figure~\ref{fig:multilayer_diagram}).

Since the layers are placed symmetrically around the $T=S+1$ line,
or more precisely around a point $(S_0, T_0)$ on this line,
the parameter $\Delta S = S^I - S^{II}$ is sufficient to determine
values of all four parameters $S^I, T^I, S^{II}, T^{II}$. Namely:
\begin{equation}
\begin{split}
& S^I = S_0 + \frac{\Delta S}{2} ,\\
& T^I = T_0 - \frac{\Delta T}{2} ,\\
& S^{II} = S_0 - \frac{\Delta S}{2} ,\\
& T^{II} = T_0 + \frac{\Delta T}{2} ,
\end{split}
\label{eqn:layers_params}
\end{equation}
where $(S_0, T_0) = (-2,-1)$ for the diagonal case and
$(S_0, T_0) = (-0.5,0.5)$ for the stag hunt case.
Note also that for $T^\beta=-S^\beta+C$, that is both cases,
we have $\Delta S = \Delta T$.
From the design of the the system it follows that there is a maximal possible
gap size $\Delta S_{max}$ above which the payoff matrices would not describe a coordination game.
In Figure~\ref{fig:multilayer_diagram} we can clearly see that $\Delta S_{max} = 4$
for the diagonal case and $\Delta S_{max} = 1$ for the stag hunt case.

We use the coordination rate $\alpha \in [0,1]$ to describe the state
of the population. When $\alpha^\beta = 1$ every player on the
layer $\beta$ chooses the strategy A, therefore layer $\beta$
is in the Pareto-optimal equilibrium.
When $\alpha^\beta = 0$ the layer also coordinates,
but on the strategy B. For $\alpha^\beta = 0.5$ both strategies
are mixed in equal amounts in the layer $\beta$. We say that the
layers are synchronised when $\alpha^I = \alpha^{II}$ and
then we use just $\alpha$ to describe both of them.
Note that synchronisation does not require coordination within
the layers and vice versa,
although they usually come together in our results.


\begin{figure}[h]
\centerline{
\includegraphics[scale=0.65]{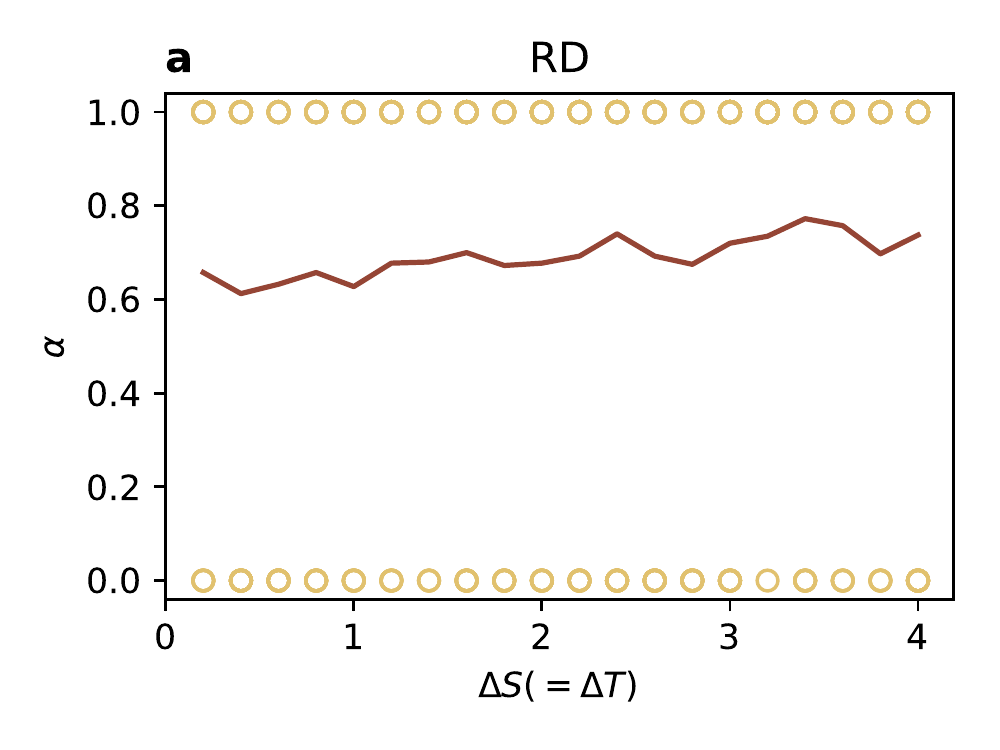}
\includegraphics[scale=0.65]{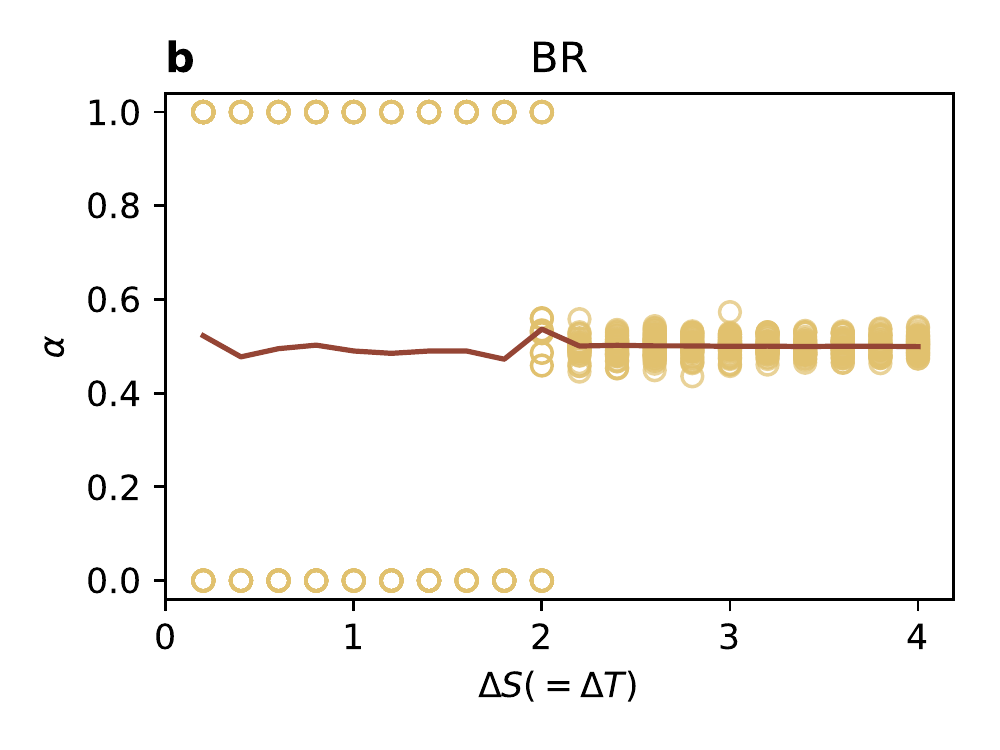}
\includegraphics[scale=0.65]{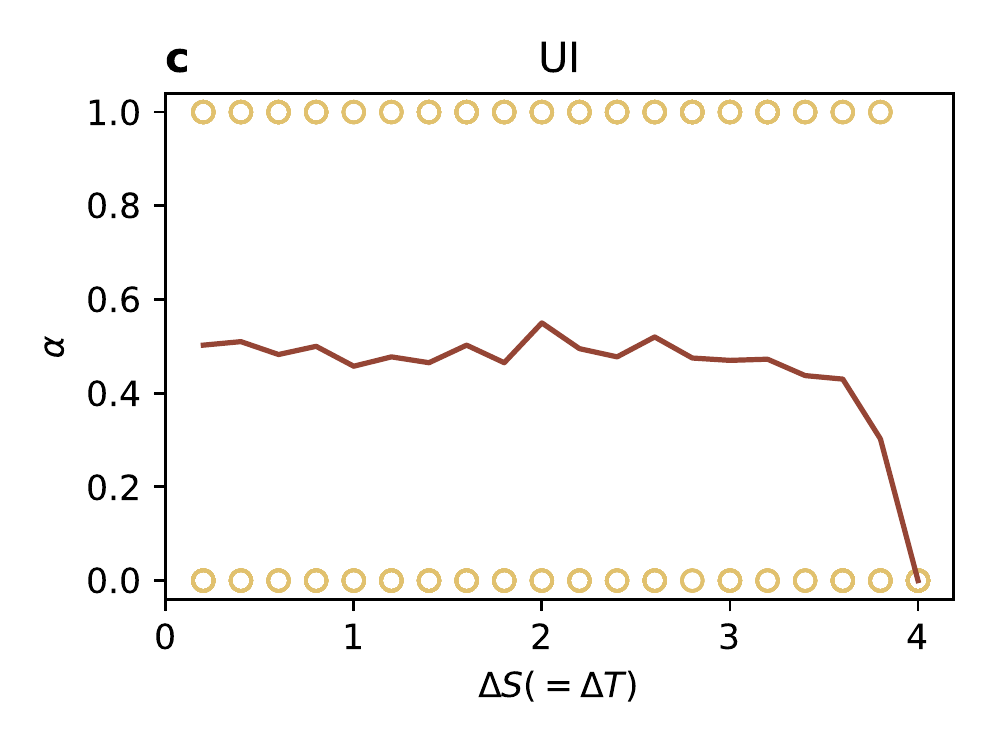}
}
\centerline{
\includegraphics[scale=0.65]{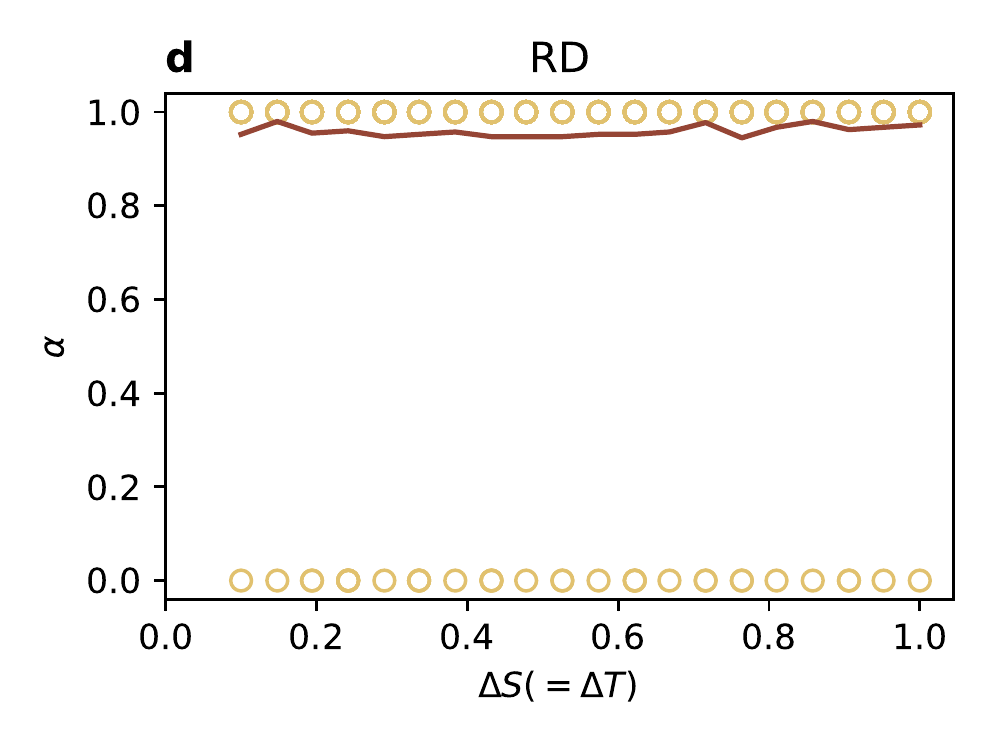}
\includegraphics[scale=0.65]{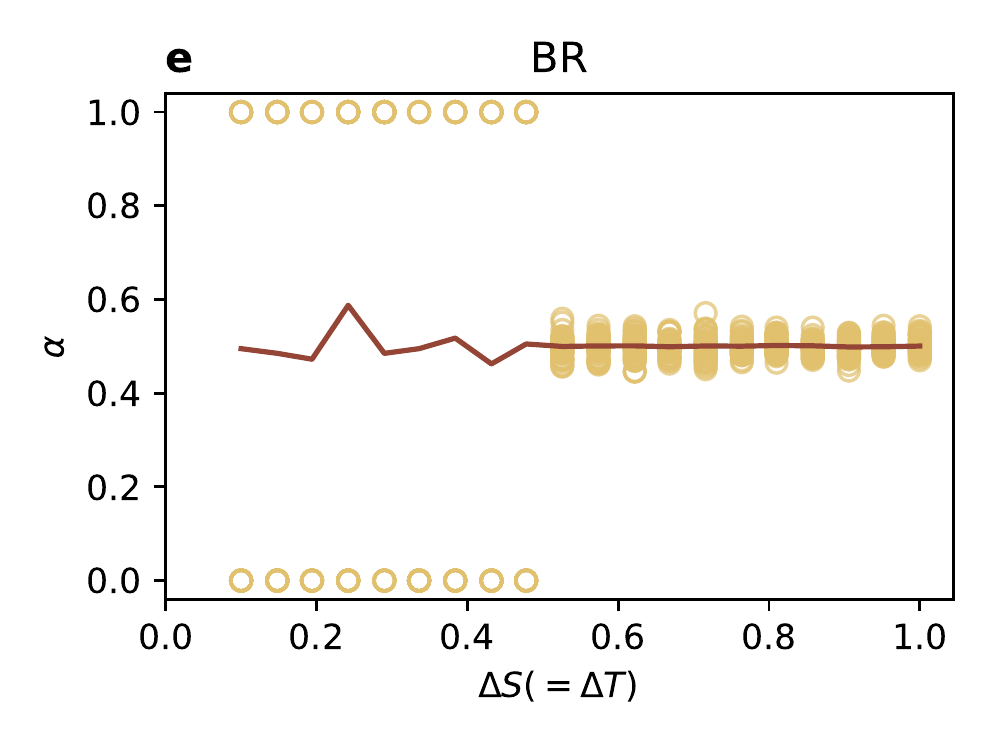}
\includegraphics[scale=0.65]{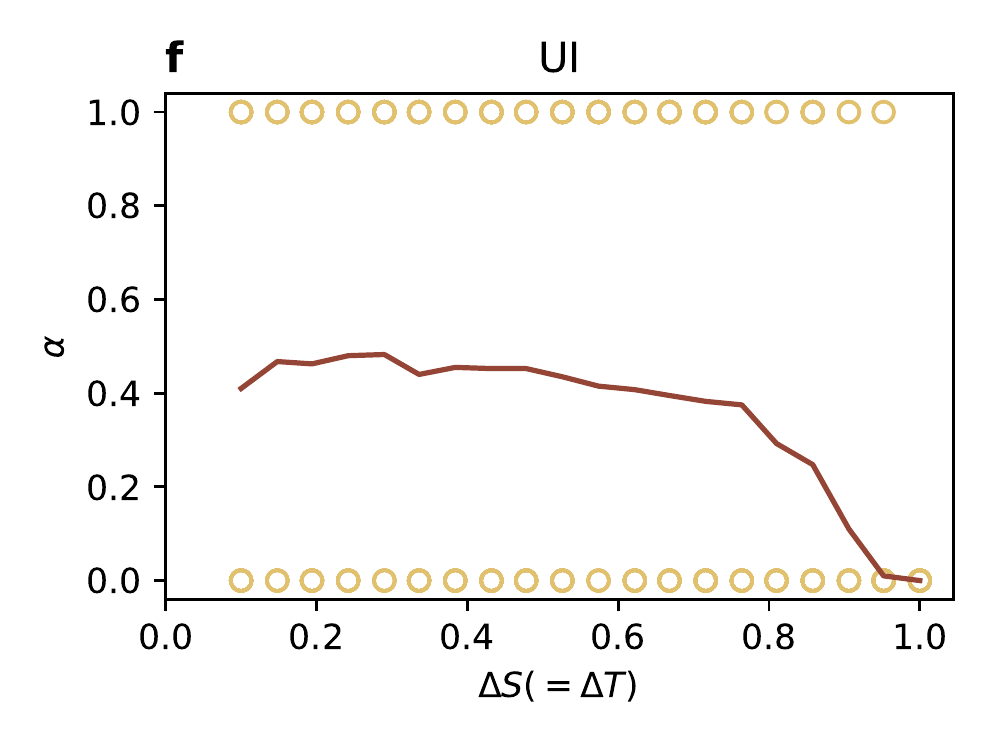}
}
\caption{
Coordination rate $\alpha=\alpha^I=\alpha^{II}$
vs gap size $\Delta S$ for full node overlap $q=1$ (the multiplex case).
The upper row (a, b, c) presents the diagonal case
and the bottom row (d, e, f) the stag hunt.
For RD and BR each layer
has $N=1000$ nodes with an intra-layer degree $k=8$, for
UI it is a complete graph with $N=500$. Each circle represents
the value of $\alpha$ (for both layers) in one of 400 realisations
and solid lines show the average values.
}
\label{fig:alpha}
\end{figure}

\section*{Results}

We study synchronisation between the layers, coordination, and
equilibrium selection under varying conditions.
For RD and BR update rules we set the connectivity
at $k=8$, since it was shown that the degree does not
change the equilibrium selection in their case
\cite{raducha2022coordination}. However, for UI the line
$T=S+1$, at which risk-dominance of strategies changes,
overlaps with the actual transition in equilibrium selection
only for a complete graph\cite{raducha2022coordination}.
Hence, we analyse the case of unconditional imitation
always with full connectivity in order to obtain true
symmetry between the layers.

The two main parameters whose influence we investigate
are the node overlap $q$ and the distance between the games
$\Delta S$ or $\Delta T$. For simplicity, we start with an analysis
of the multiplex case, i.e. full node overlap $q=1$.
In Figure~\ref{fig:alpha} we present the coordination rate
$\alpha$ for synchronised layers at $q=1$ (layers are always synchronised
at full node overlap, because all nodes have to be the same on both layers by definition).
The first thing to notice is that for the RD update rule
the system always coordinates with $\alpha = 0$ or $1$ (the circles in the figure).
In addition, the RD clearly favours the payoff-dominant strategy A
at the maximal level of multiplexity. In the diagonal case the asymmetry
is moderate with the average value of $\alpha$ between $0.6$ and $0.8$
(the solid line in the figure), but in
the stag hunt case coordination rarely happens at the strategy B
and the average value of $ \alpha$ is close to 1.

Like RD, the UI update rule always leads to full coordination
in the multiplex case with $\alpha = 0$ or $1$.
Interestingly, the UI
does not favour the strategy A. As we
can see in the figure, for small size of the gap $\Delta S$
the outcome is symmetrical with both strategies selected
half of the time. But for increasing distance between the pay-off matrices of the two layers
the system starts to coordinate more often on the strategy B,
to finally select exclusively the non Pareto-optimal equilibrium
for the maximal gap size. It has to be noted that
the maximal gap size results in payoff matrices that are on the
border between coordination games area and non-coordination games,
therefore this border point technically does not represent
a coordination game. Nonetheless, the decline in the payoff-dominant
equilibrium selection is visible already before this limit value.
This result is especially surprising, since the UI is the only update rule
that on a single-layer network can lead to the Pareto-optimal equilibrium
even though it is not risk-dominant\cite{raducha2022coordination}. However,
the requirement for the selection of a non risk-dominant equilibrium was having a sparse network and here the UI update rule is analysed
on a complete graph.

The only truly symmetric update rule is the BR which does not
reveal any preference towards one of the strategies for
full node overlap. Additionally, the diagonal case is 
identical to the stag hunt case. For gap sizes $\Delta S < \Delta S_{max} / 2$
and $q=1$ synchronised layers reach
either equilibrium with equal probability,
and for $\Delta S > \Delta S_{max} / 2$ the system does not coordinate
staying at $\alpha = 0.5$. At the transition value of
$\Delta S = \Delta S_{max} / 2$ both states -- the coordination on one of
the strategies and non-coordinated fully-mixing state -- are possible
(see Figure~\ref{fig:alpha}).

\begin{figure}[ht]
\centerline{
\includegraphics[scale=0.65]{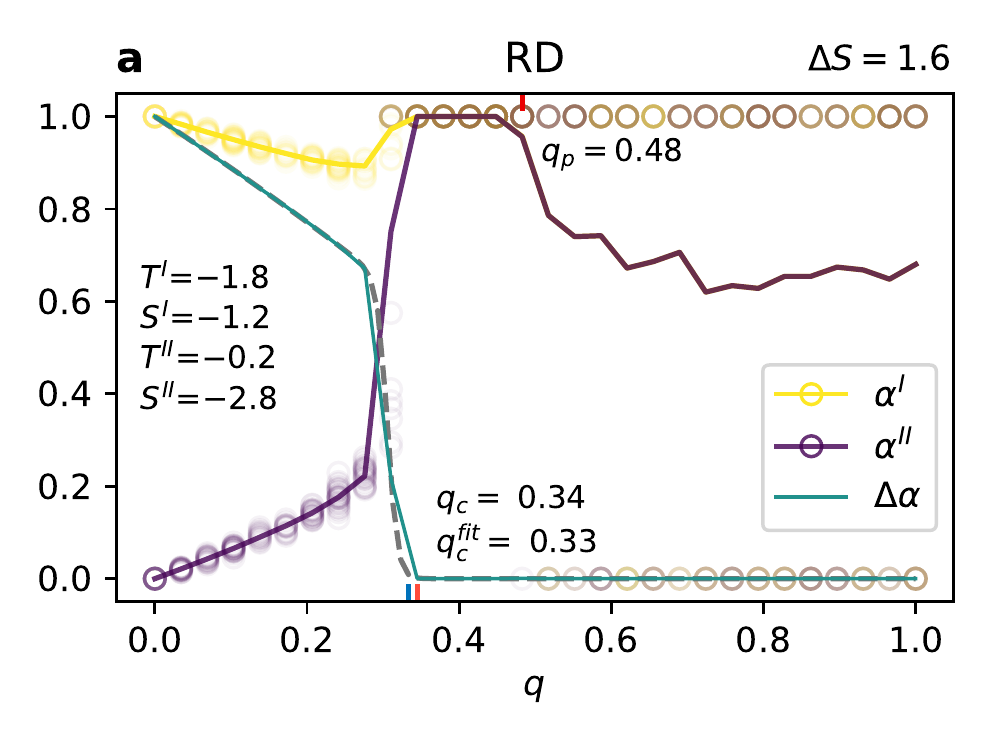}
\includegraphics[scale=0.65]{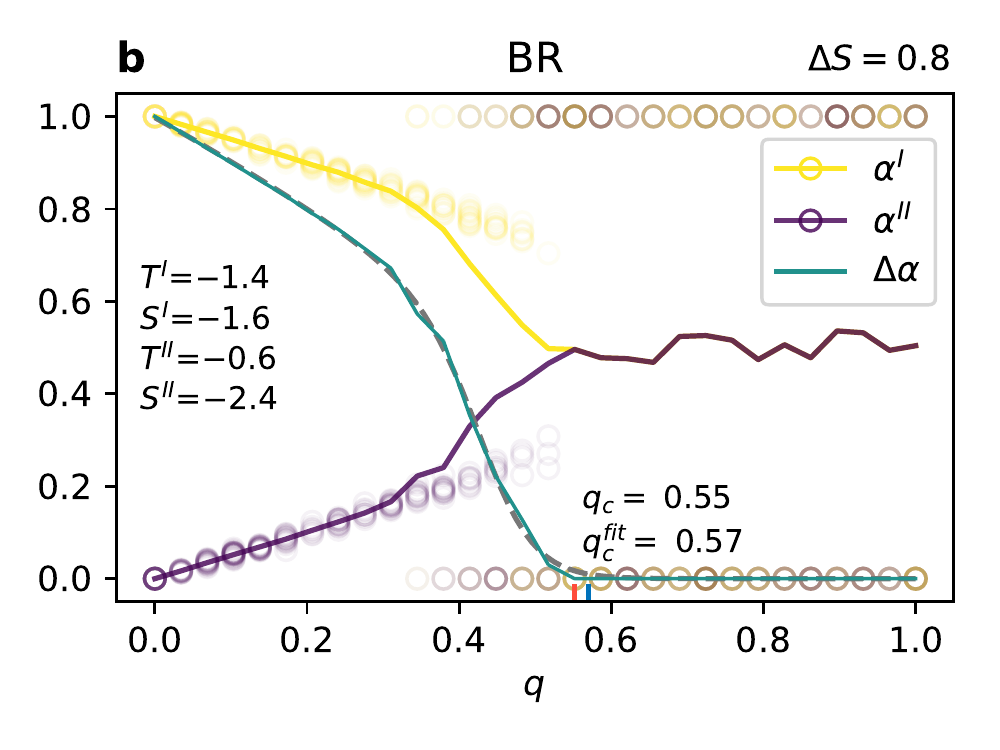}
\includegraphics[scale=0.65]{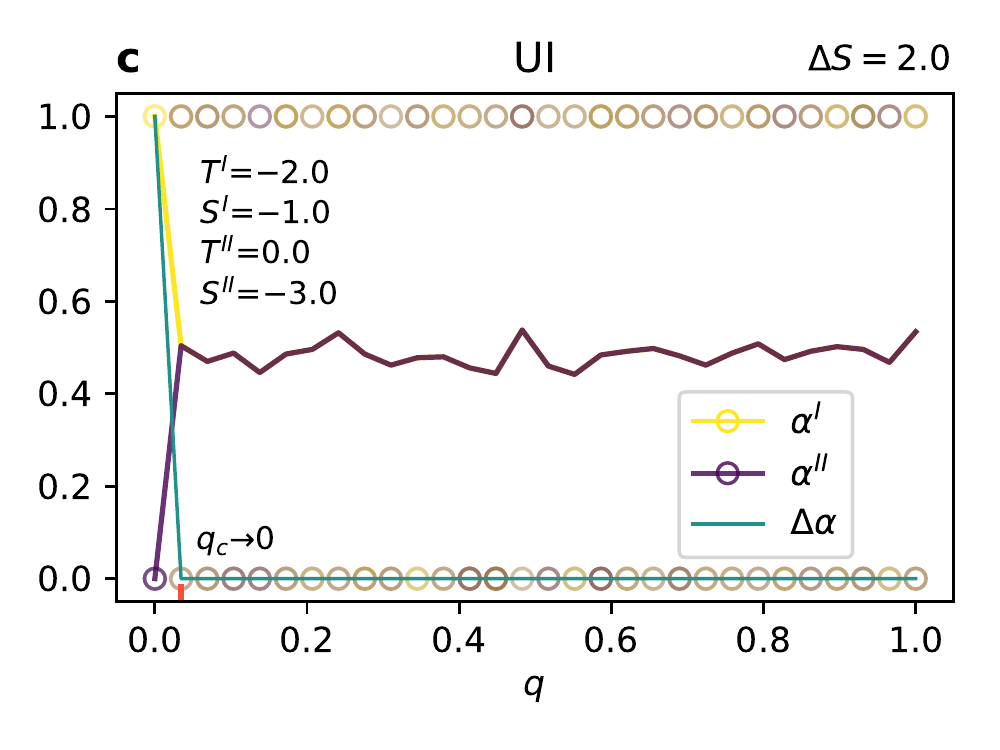}
}
\centerline{
\includegraphics[scale=0.65]{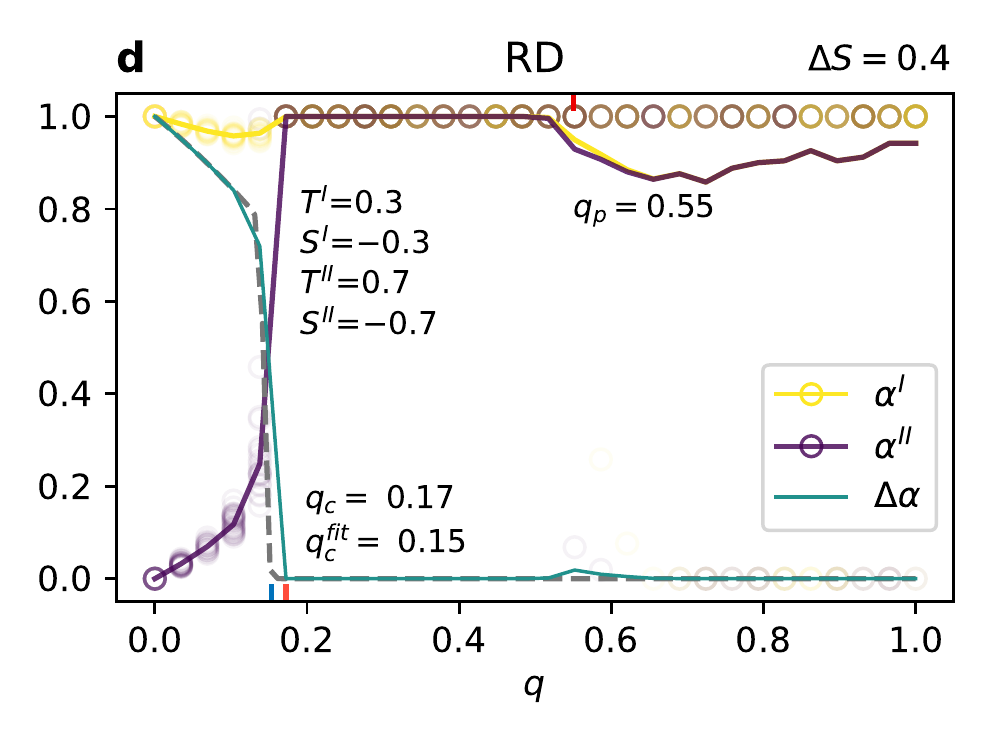}
\includegraphics[scale=0.65]{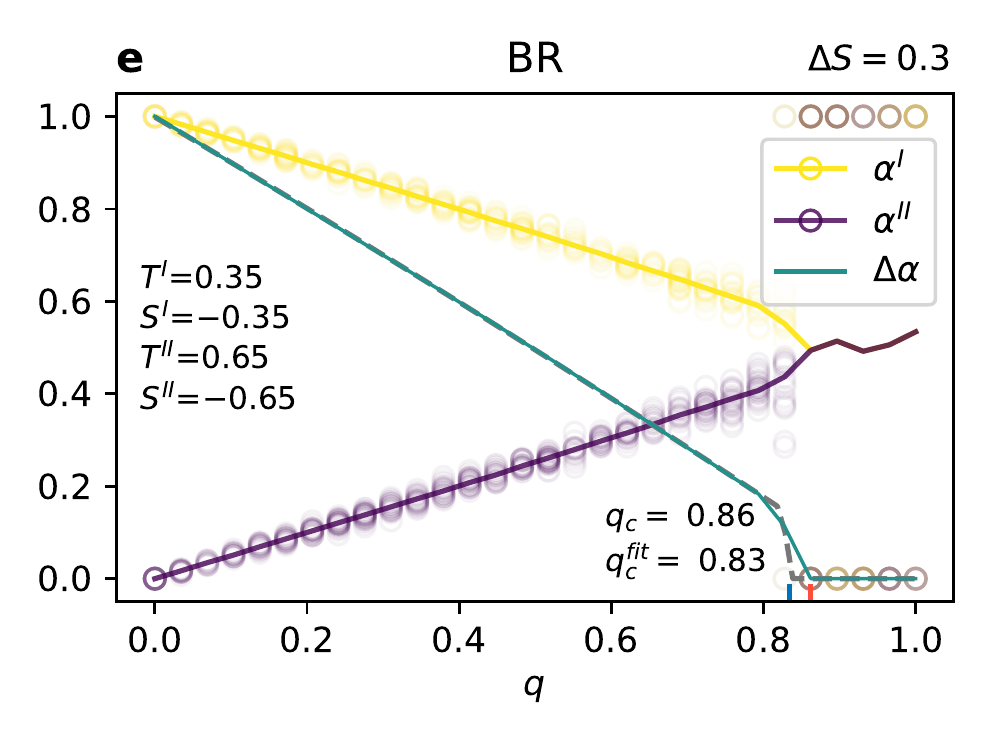}
\includegraphics[scale=0.65]{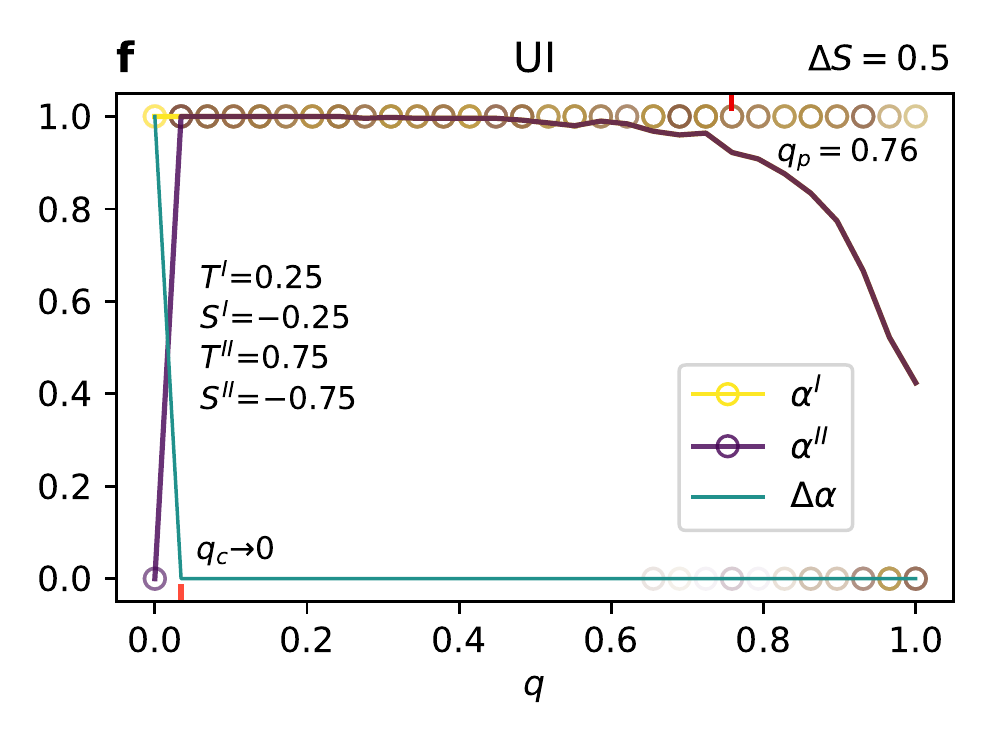}
}
\caption{
Coordination rates on layers $\alpha^{I}$, $\alpha^{II}$,
and $\Delta \alpha$ vs node overlap $q$ for exemplary
values of $\Delta S$ (see Supplementary Material for other values).
The upper row (a, b, c) presents the diagonal case
and the bottom row (d, e, f) the stag hunt.
For RD and BR each layer
has $N=1000$ nodes with an intra-layer degree $k=8$, for
UI it is a complete graph with $N=500$. Each circle represents
one of 500 realisations and solid lines show the average values.
For each realisation there is one circle for layer I (yellow) and one for layer II (purple).
Note, that when layers synchronise $\alpha^{I} = \alpha^{II}$,
 $\Delta \alpha = 0$, and both circles overlap looking like one of brownish colour,
 as well as the solid lines for $\alpha^{I}$ and $\alpha^{II}$ merge (brown).
 The dashed line in (a, b, d, e) shows a function fitted to $\Delta \alpha$.
}
\label{fig:alpha_vs_q_diag}
\end{figure}

In addition to the results showed in Figure~\ref{fig:alpha} for $q=1$, we know that
for $q=0$ each layer will obtain full coordination on its preferred strategy --
A for layer I and B for layer II \cite{raducha2022coordination}.
The middle ground between those two extreme values of $q$  must therefore contain
some kind of transition. We investigate it in Figure~\ref{fig:alpha_vs_q_diag},
where we can see how the coordination rate $\alpha$ changes at both layers
with increasing $q$. The values of $\Delta S$ in the figure were chosen
as good examples of general behaviour for each update rule, see Supplementary Material for other values.
First thing to notice is that for any update
rule and any parameter choice, but with $q=0$,
each layer converges to a different limit value of $\alpha$.
This means that both layers indeed obtain full coordination
on their preferred strategies, as expected for separate networks. Consequently,
the difference between layers is maximal $\Delta \alpha = 1$
and each network selects the risk-dominant
equilibrium. Similarly, for $q=1$ layers must fully overlap
with $\Delta \alpha = 0$, as observed in Figure~\ref{fig:alpha_vs_q_diag}, because each node is present on all
layers and the state of a shared node must be
the same across the layers.

The above considerations lead to a conclusion that there must be
a certain point $q_c \in [0,1]$ at which $\Delta \alpha$
becomes zero. In Figure~\ref{fig:alpha_vs_q_diag}
we see that the value of $q_c$ can vary for replicator
dynamics and best response update rules, but is close to zero 
for unconditional imitation. In fact,
$q_c \to 0$ for any configuration of the layers when
players update their strategies according to UI
(see Supplementary Materials for plots of different cases).
In other words, synchronisation between the layers is the strongest
for the UI update rule. One has to still bear in mind that for
UI we have considered a complete graph, while for RD and BR the networks are much
sparser with $k=8$. Nevertheless, simulations for higher degree for BR
indicate that synchronisation is weakened, not strengthened, by increasing
connectivity (see Supplementary Materials), which makes the
update rule a natural explanation of the observed differences.

Another surprising observation is that not all the results are
symmetrically placed around $\alpha = 0.5$. Both layers
have equally strong preferences towards their natural equilibria
-- payoff matrix parameters $(S^I, T^I)$ and $(S^{II}, T^{II})$
are equally distant from the transition line $T=S+1$ and
average payoffs of the games on both layers are the same.
There is no reason, in principle, why the system as a whole
should choose one equilibrium over the other. Nevertheless,
we can see that for some parameters' values with RD and UI synchronised layers
coordinate exclusively on the Pareto-optimal strategy A ($\alpha = 1$),
while it doesn't happen for the strategy B at any point
(except for $q=1$ with the maximal gap $\Delta S$ for UI, see Figure~\ref{fig:alpha}).
This symmetry breaking is especially interesting, because
it is driven by the level of multiplexity $q$ in a non-trivial way.
In examples shown in Figure~\ref{fig:alpha_vs_q_diag}, and in 
general, if the Pareto-optimal equilibrium is obtained on both layers
it happens as soon as they synchronise, i.e. at $q_c$. When increasing
the node overlap further at some point $q_p$ the synchronised state
with coordination on the strategy $B$ starts to appear and
the average value of $\alpha$ drops below 1. For $q>q_p$
synchronised layers can coordinate on either strategy,
however $\alpha>0.5$ in most cases meaning that the Pareto-optimal
equilibrium is still dominant. It is important to note that sometimes $q_c=q_p$
and the system goes directly from no synchronisation to coordination on 
either of the strategies. This is the case visible in
Figure~\ref{fig:alpha_vs_q_diag} b, c, and e, where indeed there is no pure Pareto regime.

The fully symmetrical outcome that one would expect from the symmetrical
design of the experiment is obtained solely for BR. We can
see in Figure~\ref{fig:alpha_vs_q_diag}
that there are only two types of behaviour
that the system displays with BR update rule. The first one, for $q<q_c$
is characterised by no synchronisation between the layers and each
of them following a specific level of coordination, which is
$\alpha^I = -q/2+1$ and $\alpha^{II} = q/2$. This calculation
comes from a simple assumption that all nodes that are not shared
play the dominant strategy of their layer and all shared nodes
play either strategy half of the time. Put differently,
half of the shared nodes play the strategy A and half the strategy B.
This gives a fraction $(1-q)+q/2=-q/2+1$ of nodes playing 
the strategy A on the layer I and the same fraction of nodes
playing the strategy B on layer II, so $1-(-q/2+1)$ of them playing
the strategy A. As we can see in the figure this scenario
is realised until reaching $q_c$. The second type of behaviour, for
$q>q_c$, is coordination of both synchronised layers on one of
the strategies with equal probability of choosing either of them.

\begin{figure}[ht]
\centerline{
\includegraphics[scale=0.65]{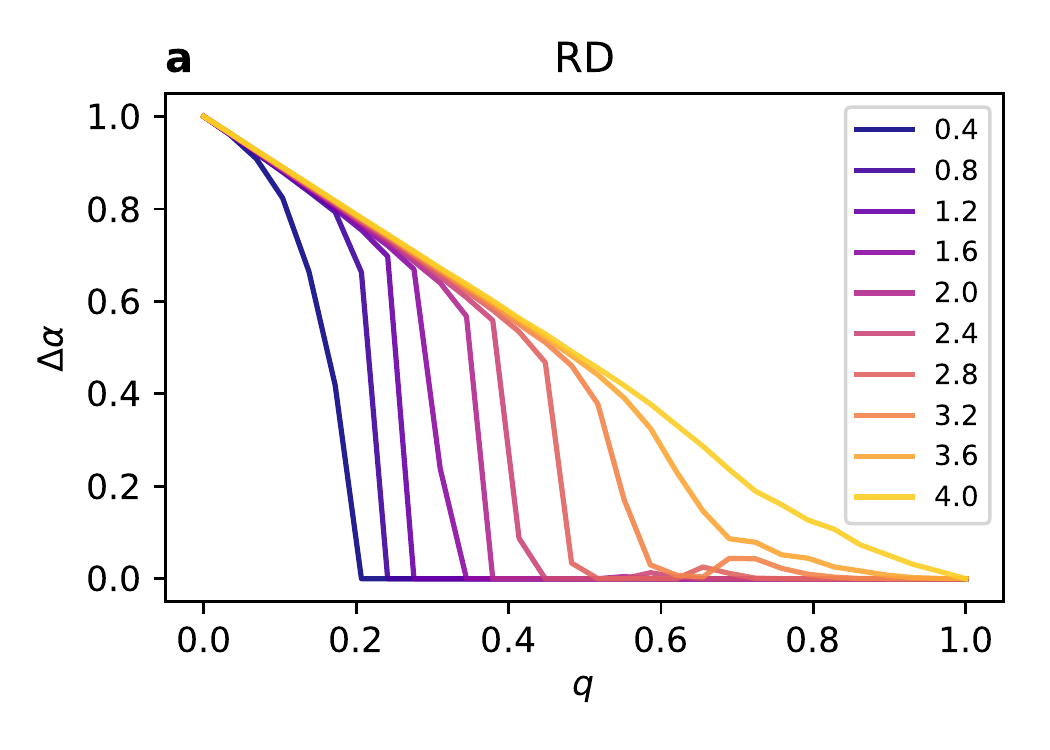}
\includegraphics[scale=0.65]{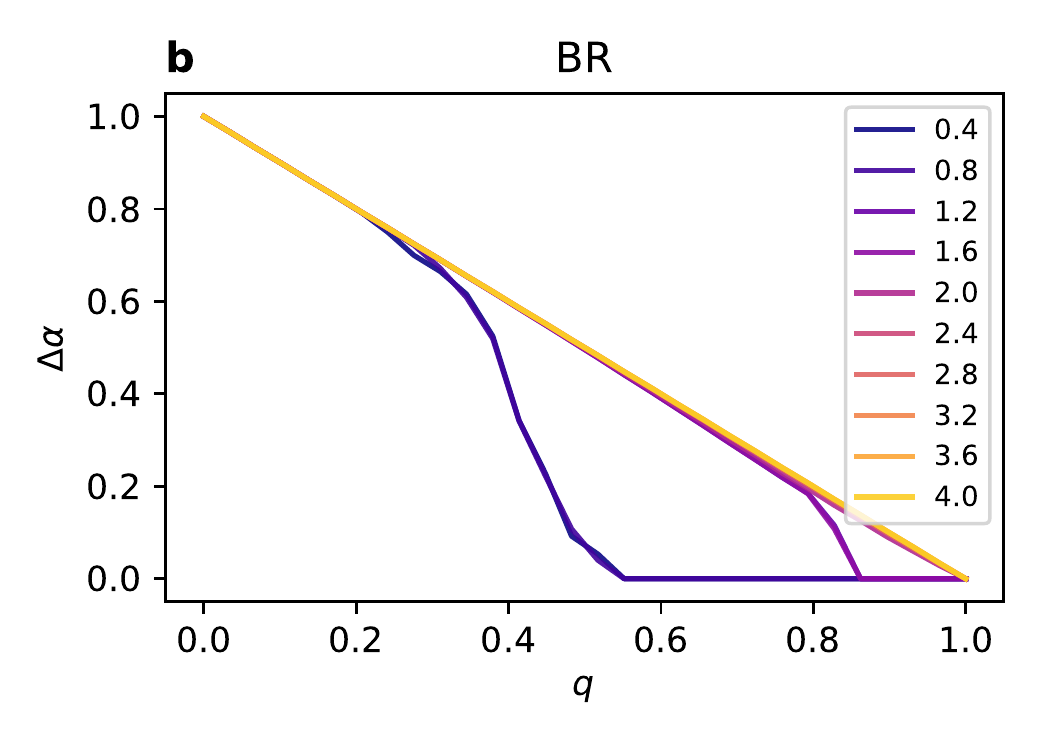}
}
\centerline{
\includegraphics[scale=0.65]{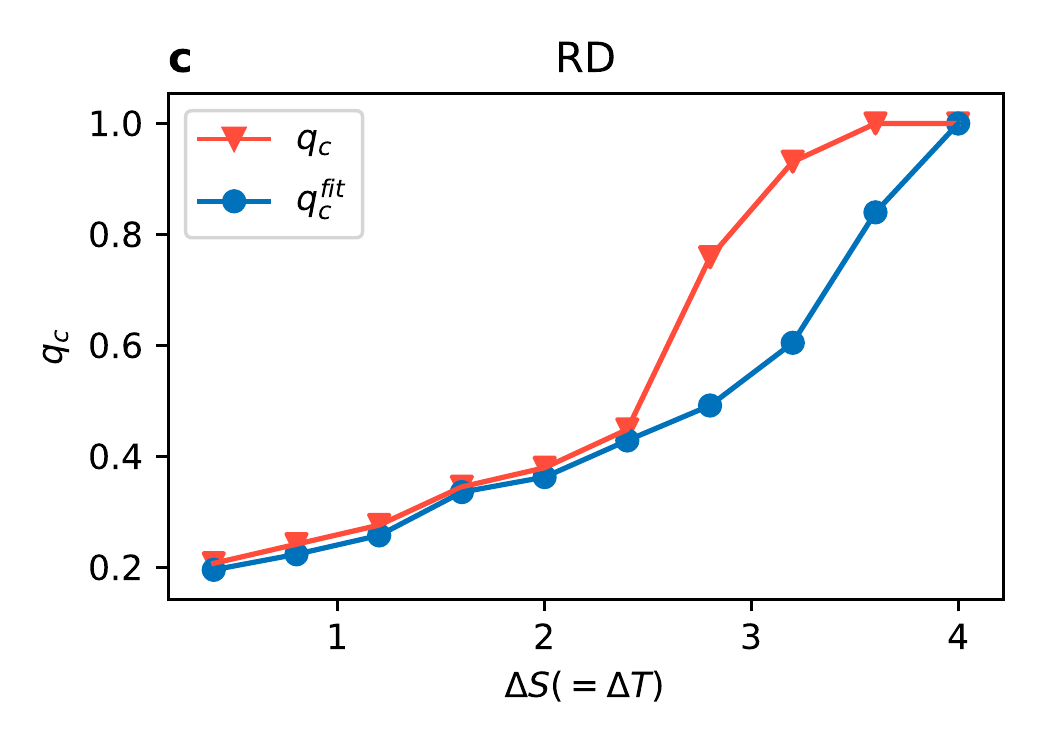}
\includegraphics[scale=0.65]{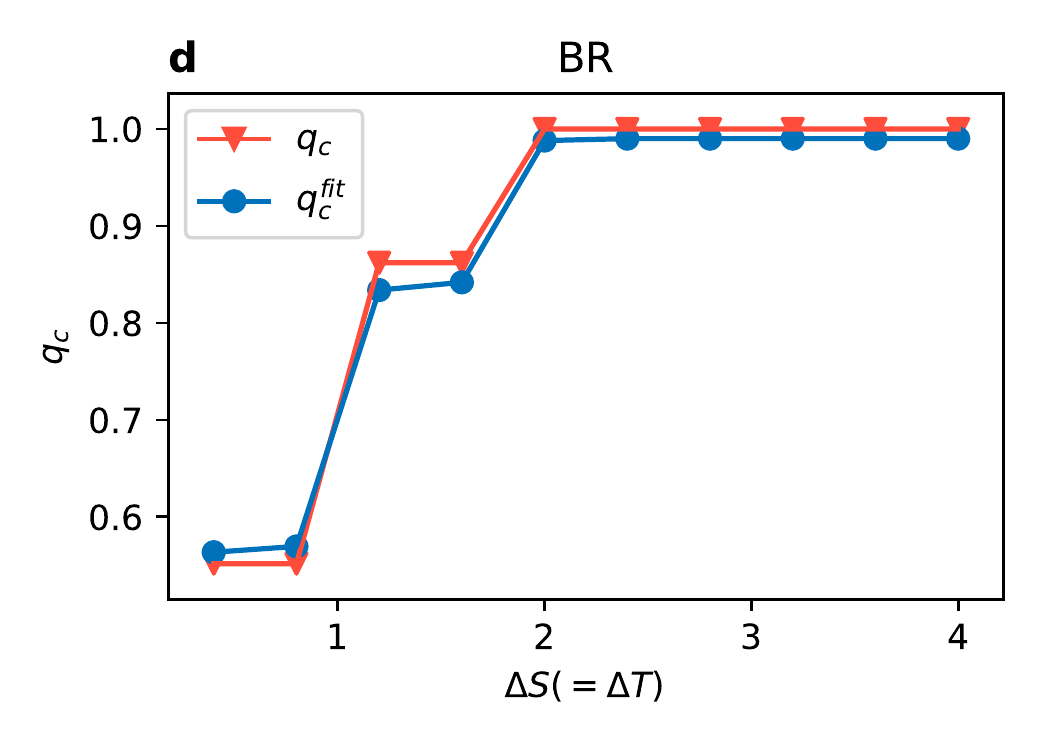}
}
\caption{
(a, b) Coordination rate difference between the layers $\Delta \alpha$
vs node overlap $q$ for increasing value of $\Delta S$ (given in the legend).
(c, d) Critical value of $q_c$ and $q_c^{fit}$ vs gap size $\Delta S$.
Results for the diagonal case,
$N=1000$ nodes on each layer with an intra-layer
degree $k=8$ averaged over 100 realisations.
}
\label{fig:alpha_scaling_diag}
\end{figure}

\begin{figure}[ht]
\centerline{
\includegraphics[scale=0.65]{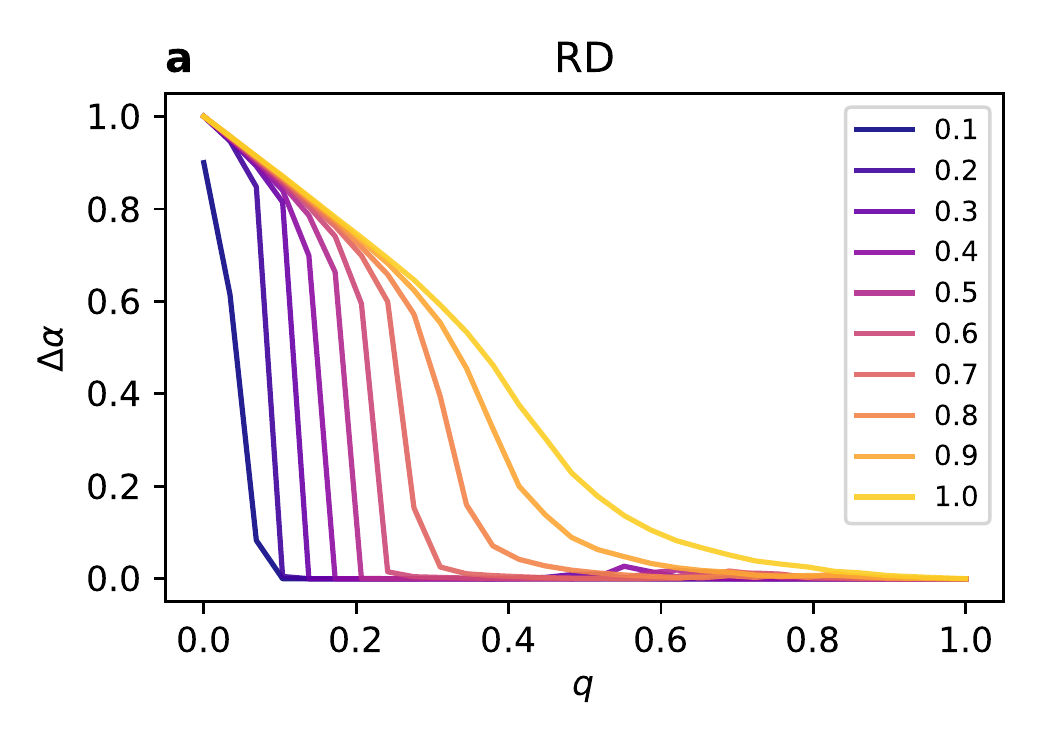}
\includegraphics[scale=0.65]{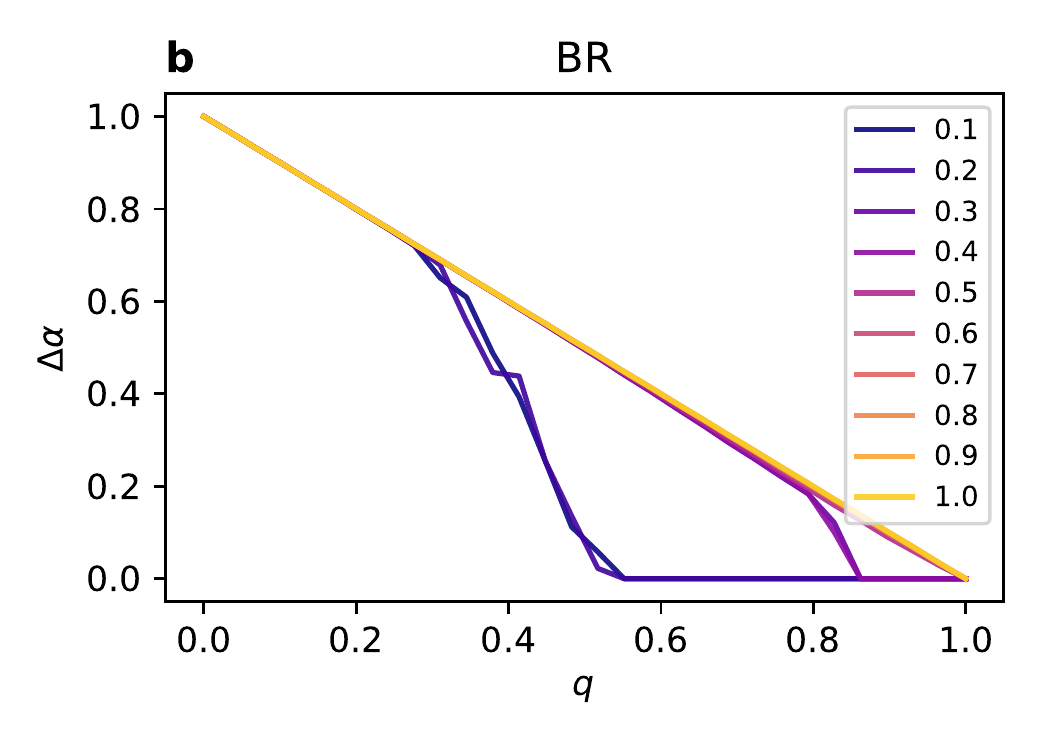}
}
\centerline{
\includegraphics[scale=0.65]{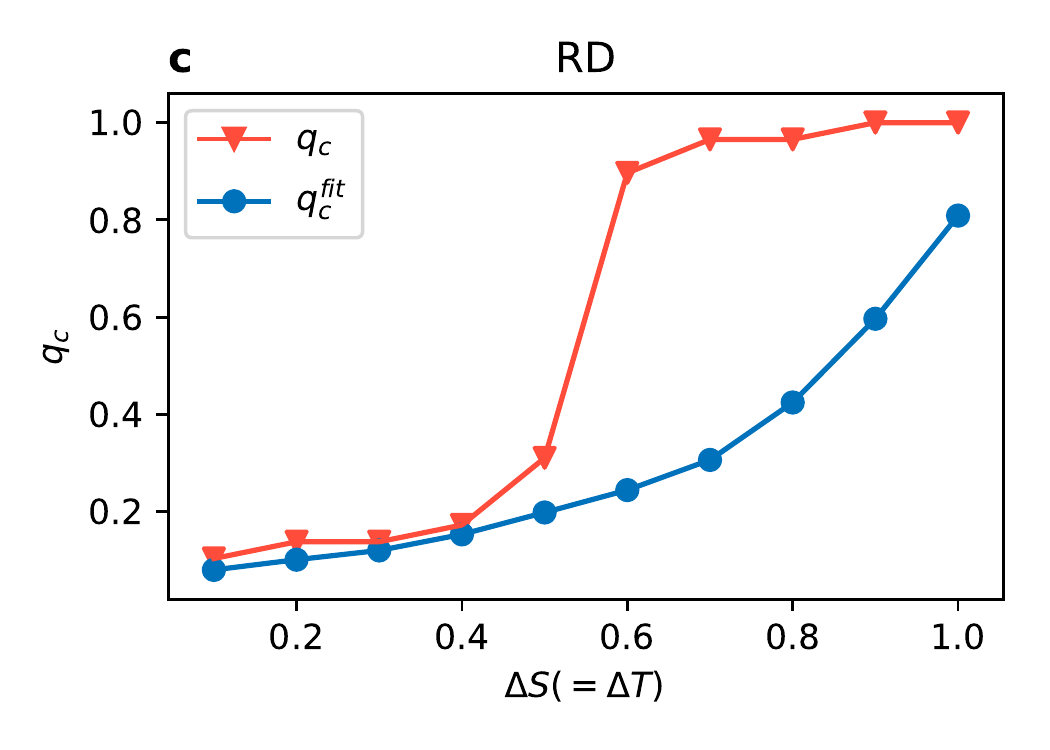}
\includegraphics[scale=0.65]{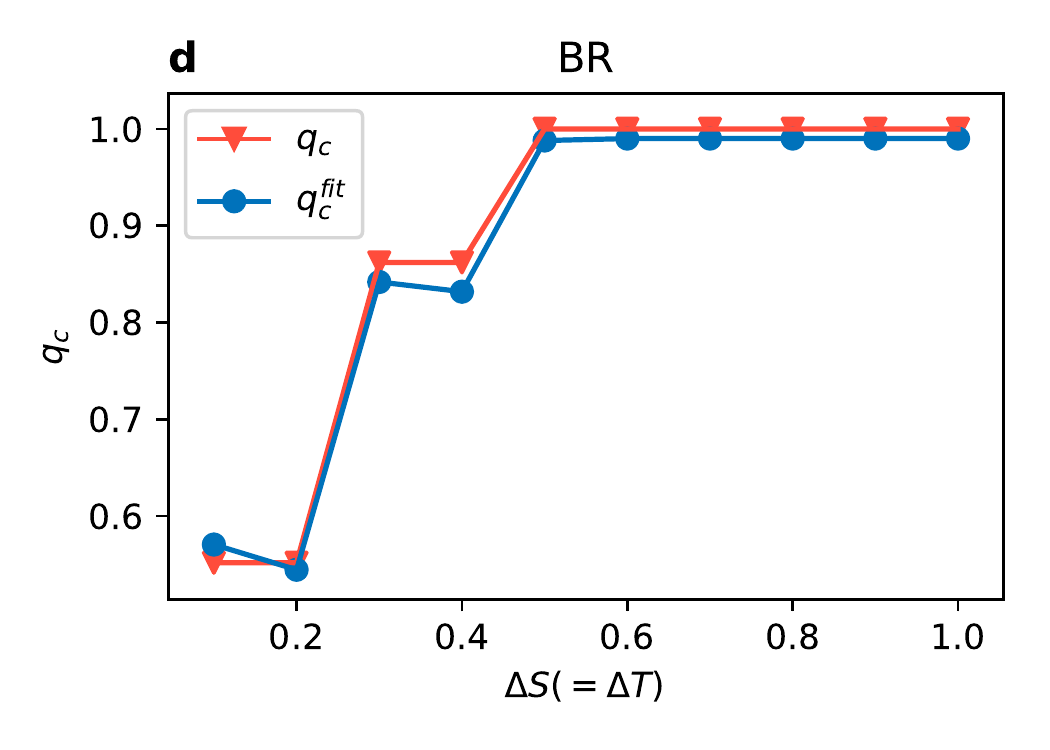}
}
\caption{
(a, b) Coordination rate difference between the layers $\Delta \alpha$
vs node overlap $q$ for increasing value of $\Delta S$ (given in the legend).
(c, d) Critical value of $q_c$ and $q_c^{fit}$ vs gap size $\Delta S$.
Results for the stag hunt case,
$N=1000$ nodes on each layer with an intra-layer
degree $k=8$ averaged over 100 realisations.
}
\label{fig:alpha_scaling_stag}
\end{figure}

The behaviour observed so far leads to a question
about the change, if any, we would observe when varying
the distance between the layers,
i.e. for different values of the gap size $\Delta S$.
In Figures~\ref{fig:alpha_scaling_diag} and~\ref{fig:alpha_scaling_stag} (a, b)
we present the dependence of $\Delta \alpha$ on the degree
of multiplexity $q$ for values of $\Delta S$ ranging from
0.4 to 4 in the diagonal case, and from 0.1 to 1 for the stag hunt.
This range essentially covers the whole width of the general coordination game
area, as presented in Figure~\ref{fig:multilayer_diagram}.
What we can see is that $\Delta \alpha$ drops to zero at higher
node overlap when increasing the gap size. More precisely,
for RD it roughly follows the line of $\Delta \alpha = -q +1$
to diverge from it at some point and eventually reach the lowest
possible value of 0. The line is followed for much longer in the
diagonal case than in the stag hunt case. For BR there is virtually
no difference between those cases and the dependence on the gap
size is slightly different. Values of $\Delta \alpha$ are the same
for gap sizes equal 0.4 and 0.8, then again for 1.2 and 1.6, and from 
$\Delta S = 2$ onwards  $\Delta \alpha = -q +1$ (these values are
for the diagonal case, for stag hunt the general picture is the same
with values rescaled by a factor of 1/4).

\begin{figure}[ht]
\centerline{
\includegraphics[scale=0.65]{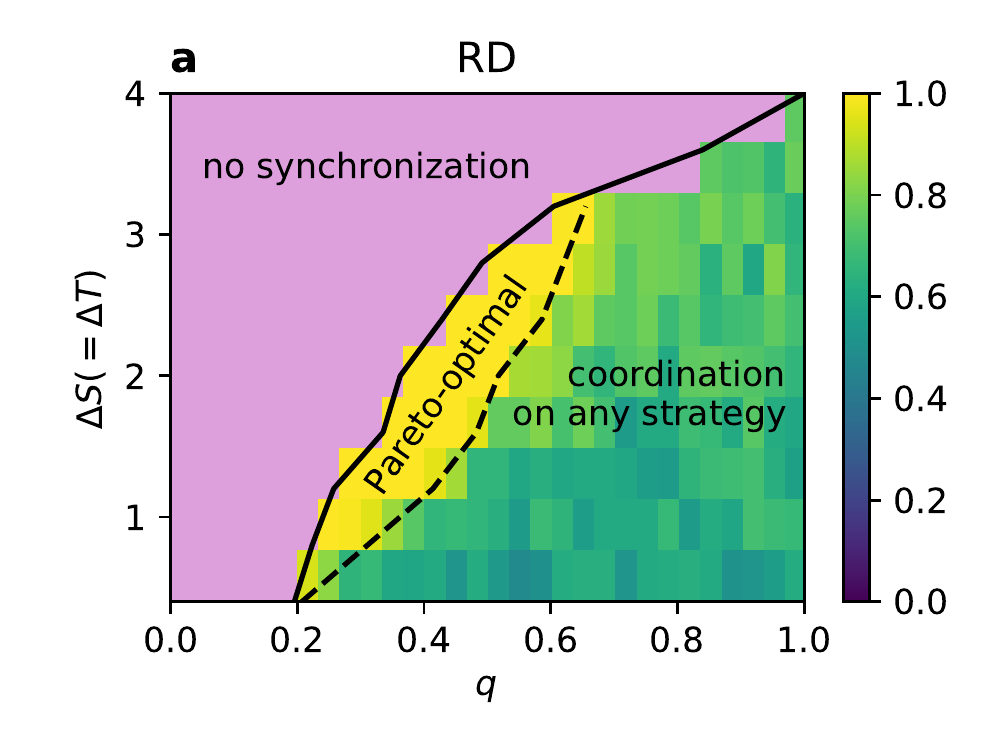}
\includegraphics[scale=0.65]{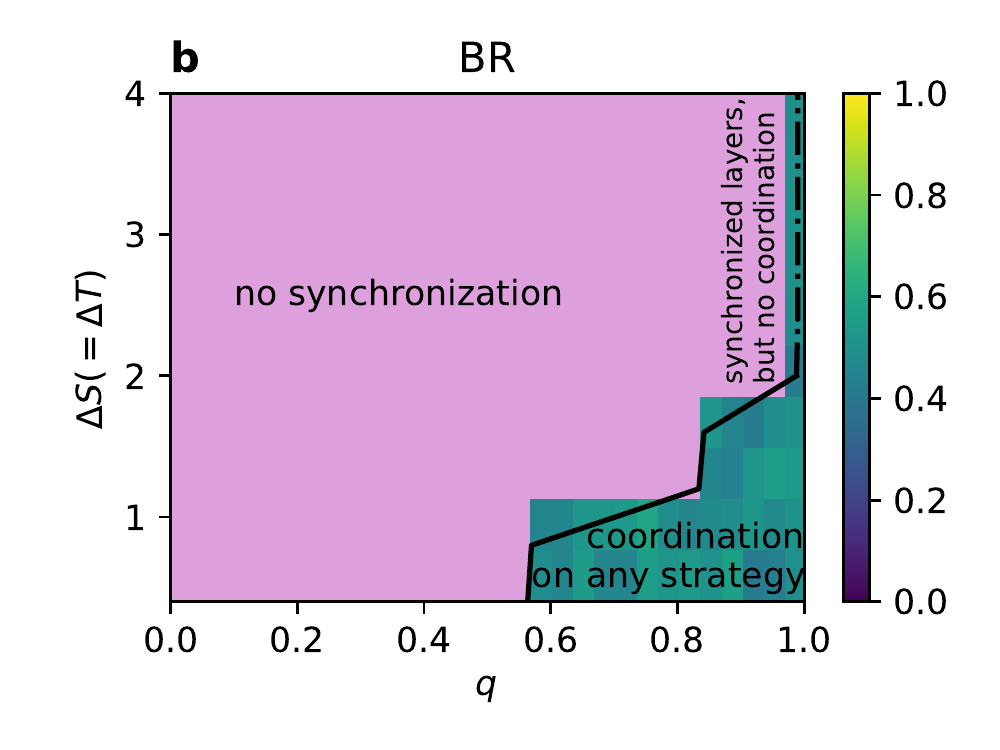}
\includegraphics[scale=0.65]{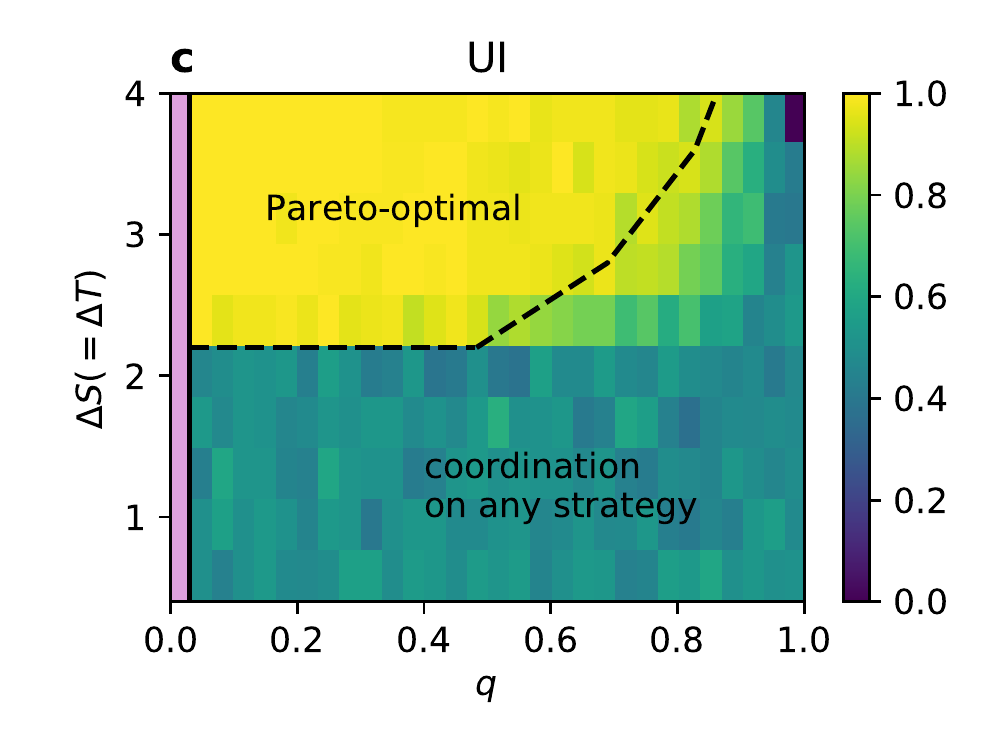}
}
\centerline{
\includegraphics[scale=0.65]{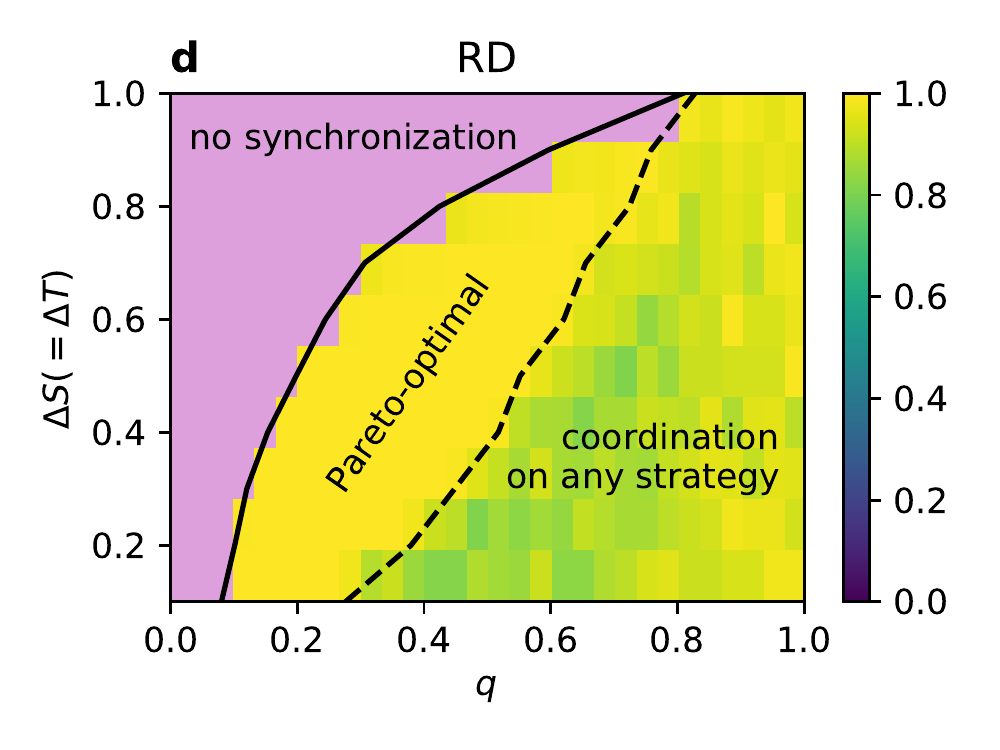}
\includegraphics[scale=0.65]{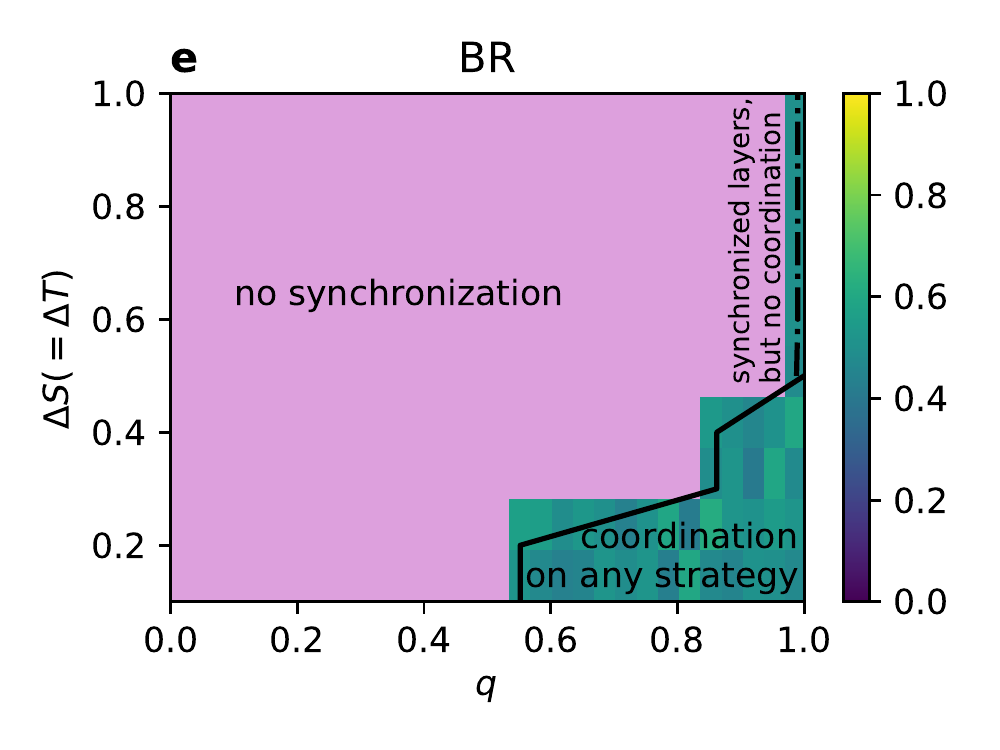}
\includegraphics[scale=0.65]{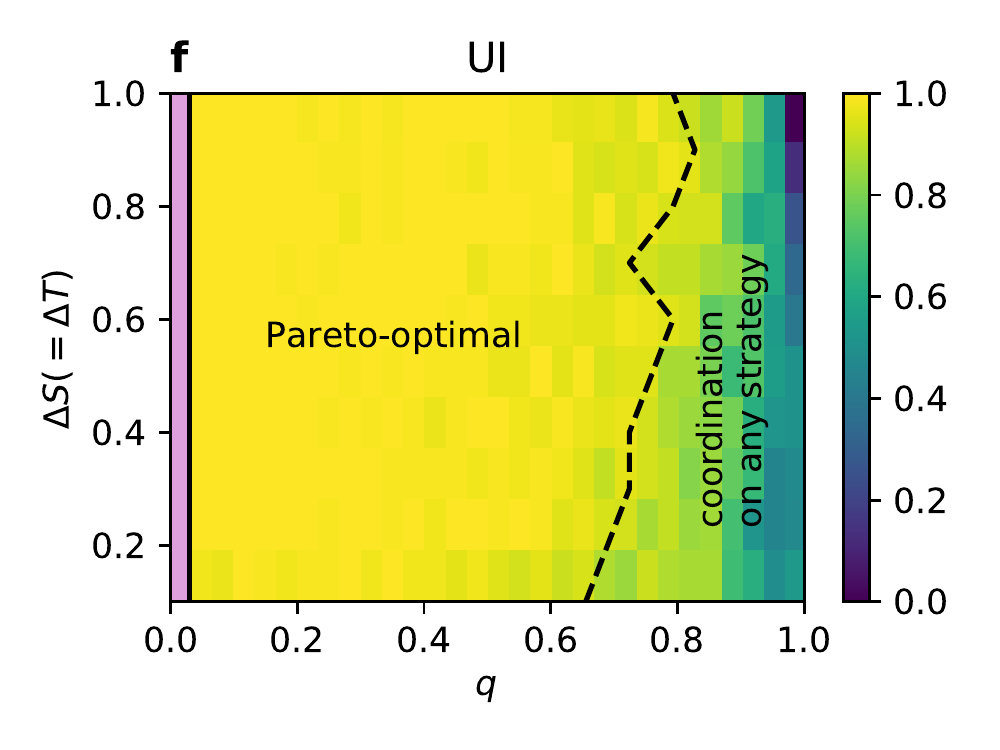}
}
\caption{
Phase diagram of coordination rate $\alpha=\alpha^I=\alpha^{II}$
in the $q$-$\Delta S$ space, for synchronised layers
for the diagonal (a, b, c) and the stag hunt (d, e, f) case.
The pink area represents the range of parameters where
synchronisation is not obtained and $\alpha^I \neq \alpha^{II}$
(for UI it happens only at $q=0$).
The solid lines show the critical value $q_c^{fit}$ and
the dashed lines $q_p$.
For RD and BR each layer
has $N=1000$ nodes with an intra-layer degree $k=8$, for
UI it is a complete graph with $N=500$. 
Results are  averaged over 100 realisations.
}
\label{fig:phase_diagram}
\end{figure}

We can clearly see that $q_c$
depends on the gap size $\Delta S$ and this dependence
is presented in Figures~\ref{fig:alpha_scaling_diag}
and~\ref{fig:alpha_scaling_stag} (c, d). We use two approaches in order to
estimate the value of $q_c$. The first one is simply taking the
lowest value of $q$ at which $\Delta \alpha$ is equal 0
for the first time. This approach, however, is prone to numerical
noise and a tiny divergence from 0 will result in a change of the value.
To obtain the second one we fit a parabola with an exponential
cutoff to the function $\Delta \alpha (q)$ (dashed line in Figure~\ref{fig:alpha_vs_q_diag}) and we take the first
value of $q$ at which $\Delta \alpha < 0.01$ as $q_c^{fit}$.
As we can see in the plots, it does not make a real difference
for BR, but can give different results for RD for higher values
of $\Delta S$. Regardless the approach, $q_c$ changes from approximately
0.2 up to 1 for RD in the diagonal case (for the stag hunt values are slightly
lower), and from 0.5 to 1 for BR with no visible difference between
the diagonal and the stag hunt case.
We similarly estimate the value of $q_p$, however without fitting a function,
because the behaviour of  $\alpha$ for synchronised layers
is more complex than the one of $\Delta \alpha$.
We take as an approximation of $q_p$ the first value of $q$
after synchronisation for which the coordination rate
$\alpha$ drops below 0.95  (dashed lines in Figure~\ref{fig:phase_diagram}).

In summary, for any gap size $\Delta S$ (or $\Delta T$) between the layers
at $q=0$ there is no synchronisation and each layer gravitates
towards its preferred equilibrium. Then, at $q=q_c$ layers start
to synchronise. For RD and UI synchronised layers coordinate 
on the Pareto-optimal strategy for $q_c<q<q_p$ and for $q>q_p$
they coordinate on either of the strategies. For some values
of $\Delta S$, however, as well as for BR in general, $q_p$
overlaps with $q_c$ and the system goes from unsynchronised state
straight into coordination on any strategy, without the phase of
pure Pareto-optimal equilibrium. We illustrate all these results with phase diagrams
in the $q$-$\Delta S$ space in Figure~\ref{fig:phase_diagram}.
Additionally, there are two 
update-rule-specific phenomena.
For UI at the maximal
gap between the layers ($\Delta S_{max} = 4$ for the diagonal case and
$\Delta S_{max} = 1$ for the stag hunt) and for $q=1$ synchronised layers
coordinate only on the strategy B just at this point.
And for BR for $\Delta S > \Delta S_{max} / 2$ at full node overlap
when the layers get synchronised they do not reach coordination.
Instead they both end up in a fully mixing state with
$\alpha^I=\alpha^{II}=0.5$ (see panels b and e of
Figure~\ref{fig:alpha} and of Figure~\ref{fig:phase_diagram}).

We can also see from
Figure~\ref{fig:phase_diagram}
that an increase in the absolute values of payoffs $S^\beta$ and $T^\beta$
on both layers,
i.e. a shift from the diagonal to the stag hunt case, significantly
enlarges the relative area of Pareto-optimal equilibrium for RD and UI.
It does not, however, change the relative size of the no-synchronisation
phase and it seems not to influence the best response dynamics at all.
One explanation of the enlargement of the Pareto-optimal phase,
at least for RD, could be the fact that in the stag hunt case
the layers are closer to each other -- the gap size $\Delta S$
(and $\Delta T$) is 4 times smaller on average. Games being more
similar and closer to the transition line could justify
why it is easier for layer I to shift layer II into
its preferred equilibrium on the strategy A.
Nevertheless, for UI in the diagonal case there is a minimal value
$\Delta S \approx 2$ below which the Pareto-optimal phase
does not exist at all, hence here the proximity of layers can not be the 
explanation of synchronisation in the payoff-dominant equilibrium.
Moreover, there is an optimal size of the gap $\Delta S$
for which the Pareto-optimal phase is the widest.
For UI it is approximately the maximal gap $\Delta S_{max}$
and for RD it is one of the middle values, but certainly not
the smallest gap. These considerations lead us to a conclusion
that synchronisation and equilibrium selection in
coordination games on multilayer networks are very complex
phenomena where obtaining the most advantageous outcome
requires accurate parameter selection.

\section*{Discussion}

We investigated synchronisation between layers
and equilibrium selection in the general coordination game
on a multilayer network. The game played on each layer
is described by a different payoff matrix, but both games are
equally distant from the risk-dominance transition line $T=S+1$.
The layers are connected  by $Nq$ inter-links,
where the parameter $q$ is the node overlap or degree of multiplexity. We studied the impact of
of the value of  $q$ and the gap $\Delta S$ between the layers
for three update rules: the replicator dynamics, the best response,
and the unconditional imitation.

The most prominent outcome is the symmetry breaking
in equilibrium selection. In neither of the cases,
diagonal and stag hunt, there is a difference in
average payoffs of games played on the layers. The strategies
preferred by each layer are equally risk-dominant, i.e.
the distance from the transition line $T=S+1$ is the same.
The only difference, of course, is that the strategy A
gives the highest possible payoff, hence it's the most
profitable one. A common-sense approach would lead us to
believe that the payoff-dominant strategy A should be
naturally promoted by the population. This is however not
the case on single-layer networks, where the risk-dominant
strategy is always selected in the range of connectivities
that we considered \cite{raducha2022coordination}.
In our multilayer model, which strategy is risk-dominant 
depends on the layer, but coordination on the strategy A prevails
in most of the parameters space or is at least favoured on average.
It is therefore clear that the multilayer structure
enhances the Pareto-optimal outcome and it does so in 
a complex manner.

We identified three main phases depending on the node overlap $q$
and the gap size $\Delta S$. The first one for lower values of $q$
is a no-synchronisation phase with $\alpha^I \neq \alpha^{II}$.
Each layer obtains a certain level of
coordination close to its preferred equilibrium. The second phase
begins when $\Delta \alpha$ drops to zero, i.e. at $q_c$.
Here, layers are synchronised and fully coordinate on the
Pareto-optimal strategy A. Finally, the third phase appears
for a higher node overlap $q>q_p$. In this phase layers are also
synchronised and they also coordinate, but not always on
the strategy A -- either equilibrium is possible, although
depending on the parameters one of them might be preferred on
average. In some cases $q_c = q_p$ and the second phase does not
appear.

The Pareto-optimal phase is not a mere effect of high node overlap
between layers or low gap size. It has a more complex shape that
depends on both parameters and on the update rule. For BR the
Pareto-optimal phase does not exist at all. For RD it is
placed, surprisingly, in the middle rage of the node overlap $q$,
but its position and width depend also on $\Delta S$. Neither too
low nor too high degree of multiplexity helps in achieving the optimal
equilibrium, and the same is true for the gap size.
Nevertheless, the value of $q_c$ grows with increasing distance $\Delta S$.
For UI the Pareto-optimal phase might not even exist
for lower values of $\Delta S$. If the phase exists, however,
it appears already for any $q>0$, as the synchronisation is
much faster for UI.

Our work contributes to the understanding of 
equilibrium selection in coordination games, bringing in the general context of 
multilayer networks. Since many socio-technical
systems have multiple environments where people can interact,
the application of layered structures in their modelling is 
a natural step forward. As we showed, this approach can be
highly relevant in analysis of coordination dilemmas, because
it leads to non-trivial new effects that have not been observed
in single-layer networks.


\section*{Methods}

We run numerical simulations of the general coordination game
defined by the payoff matrix~(\ref{eqn:matrix_most_general})
on a multilayer graph.
Agents are placed on two networks of $N$ nodes forming
two layers of the multilayer network. Each layer is a random regular graph
with a degree $k$, generated using \verb,K_Regular, algorithm form the
\textit{igraph} python package \cite{csardi2006igraph,igraph}.
The coupling between layers can be adjusted using
two parameters: node overlap $q$ and edge overlap. As we didn't observe
any influence of varying edge overlap on the results we maintain
a perfect edge overlap, i.e. both layers have exactly
the same structure of connections.
The node overlap $q$ takes values from 0 to 1, defining
the fraction of nodes connected (or shared) between both layers.
If two nodes are shared, their state has to be the same
on both layers at all times.
In other words, it's the same node present on both layers.
For $q=0$ there is no connection between
the layers and their dynamics are fully separated,
for $q=1$ it's effectively a single-layer network
with each game played half of the time.

The game played on each layer
is described by different values of $S^\beta$ and
$T^\beta$ parameters of the payoff matrix, given in
equation~(\ref{eqn:layers_params}).
We use an asynchronous algorithm where at the beginning of each time step
a layer is randomly selected with equal probability for both layers. 
Then, the update is performed on the chosen layer
as for a single-layer network and according
to the game played on the layer.
First, a random node is chosen with equal probability for all nodes
on the layer. We call it the active or focal node.
The active node then plays the game with all its $k$ neighbours
on the layer and receives a given payoff, which is saved.
Finally, the strategy of the active node is updated according
to one of the following three update rules:
\begin{itemize}
    \item {the Replicator Dynamics (RD)} (aka replicator rule, or proportional imitation rule) -- the active node compares the payoff with a random neighbour on the layer and copies its strategy with probability $p=(\mathrm{payoff~diff.})/\phi$, if the neighbour's payoff is bigger. Normalisation $\phi$ is the largest possible payoff difference allowed by the payoff matrix and network structure and it sets the probability $p$ within $[0,1]$ range,
    \item {the myopic Best Response (BR)} -- the active node chooses the best strategy given the current strategies of the neighbours on the layer, i.e. it compares all payoffs it would obtain playing each possible strategy against the current strategies of the neighbours and chooses the strategy resulting in the largest payoff,
    \item {the Unconditional Imitation (UI)} -- the active node copies the strategy of the most successful neighbour on the layer, i.e. the one with the highest payoff, if its payoff is bigger.
\end{itemize}
At the end, the state of the focal node is copied onto
the other layer, if the updated node is connected (shared) between the layers.
More precisely, the new strategy selected by the node and
the last payoff are copied. The simulation
runs until a stationary state is reached,
or a frozen configuration is obtained on all layers.

\bibliography{bib}


\section*{Acknowledgements}

We acknowledge financial support  from MCIN/AEI/10.13039/501100011033 and the Fondo Europeo de Desarrollo Regional (FEDER, UE) through project APASOS (PID2021-122256NB-C21) and the María de Maeztu Program for units of Excellence in R\&D, grant CEX2021-001164-M.


\section*{Author contributions statement}

M.S.M. and T.R. conceived and designed the research, T.R. conducted the simulations and wrote the initial manuscript, M.S.M. and T.R. analysed the results and reviewed the manuscript. 

\section*{Competing interests}

The authors declare no competing interests.

\section*{Additional information}

\textbf{Supplementary information} is available for this paper at

\end{document}


{
\small
\maketitle}
\thispagestyle{empty}

In this supplementary material we provide additional plots of coordination rate $\alpha$
for $q=1$ (as in Figure~2 from the main manuscript), plots of coordination rate
$\alpha^I$ and $\alpha^{II}$ on each layer for varying $q$ (as in Figure~3 from the main manuscript),
plots of the dependence of $q_c$ on the gap $\Delta S$ or $\Delta T$ (as in Figure~4 from the main manuscript),
and phase diagrams showing the dependence of synchronisation and coordination on $q$ and $\Delta S$ or $\Delta T$
(as in Figure~6 from the main manuscript). In addition to the cases considered in the main text
we show results for \textit{horizontal} and \textit{vertical} cases, which are depicted in
Supplementary Figure~S1.

For the horizontal case all points lay on the line $T^\beta=-1$ and for the vertical case
on $S^\beta=-2$, $\beta \in \{\textrm{I, II}\}$.
Since the layers are placed symmetrically around the $T=S+1$ line,
or more precisely around a point $(S_0, T_0)$ at this line,
the parameter $\Delta S = S^I - S^{II}$ or $\Delta T = T^I - T^{II}$ is sufficient to determine
values of all four parameters $S^I, T^I, S^{II}, T^{II}$. Namely:
\begin{equation}
\begin{split}
& S^I_{h} = S_0 + \frac{\Delta S}{2} ,\\
& T^I_{h} = - 1,\\
& S^{II}_{h} = S_0 - \frac{\Delta S}{2} ,\\
& T^{II}_{h} = -1 ,
\end{split}
\label{eqn:horizon_params}
\end{equation}
for the horizontal case and:
\begin{equation}
\begin{split}
& S^I_{v} = -2 ,\\
& T^I_{v} = T_0 - \frac{\Delta T}{2} ,\\
& S^{II}_{v} = -2 ,\\
& T^{II}_{v} = T_0 + \frac{\Delta T}{2} ,
\end{split}
\label{eqn:vert_params}
\end{equation}
\newpage \thispagestyle{empty} \noindent
for the vertical case, where $(S_0, T_0) = (-2,-1)$.
Note, that horizontal and vertical cases are not fully symmetrical in respect
to the games played on the layers. In the horizontal case the game played on
layer I has always higher average payoff and in the vertical case the average payoff
is bigger on layer II.

It is worth mentioning that results for BR for higher connectivities indicate that
bigger degree may destroy synchronisation (Supplementary Figures~S8, S10, and S12).
Synchronisation is also weakened in the horizontal case for RD when changing the connectivity
from $k=8$ to a complete graph (Supplementary Figure~S5).
Additionally, it seems to be detrimental for the Pareto-optimal
equilibrium.
It is especially interesting, since the Pareto-optimal phase is abundant
for UI on a complete graph, except for the vertical case in which
the layer II has higher average payoff (Supplementary Figures~S14-S17).
Finally, the Pareto-optimal phase and synchronisation region are bigger
in the vertical case than in the horizontal case for RD
(Supplementary Figures~S4 and S6). This is especially surprising,
because the vertical case is biased towards layer II, as mentioned before, and
the horizontal case favours layer I (in terms of the average payoff).
The horizontal case results in a bigger Pareto-optimal phase
for UI, but surprisingly the effect is inverted for RD.

\vspace{1cm}
\begin{figure}[th!]
\centering
    \includegraphics[scale=1.1]{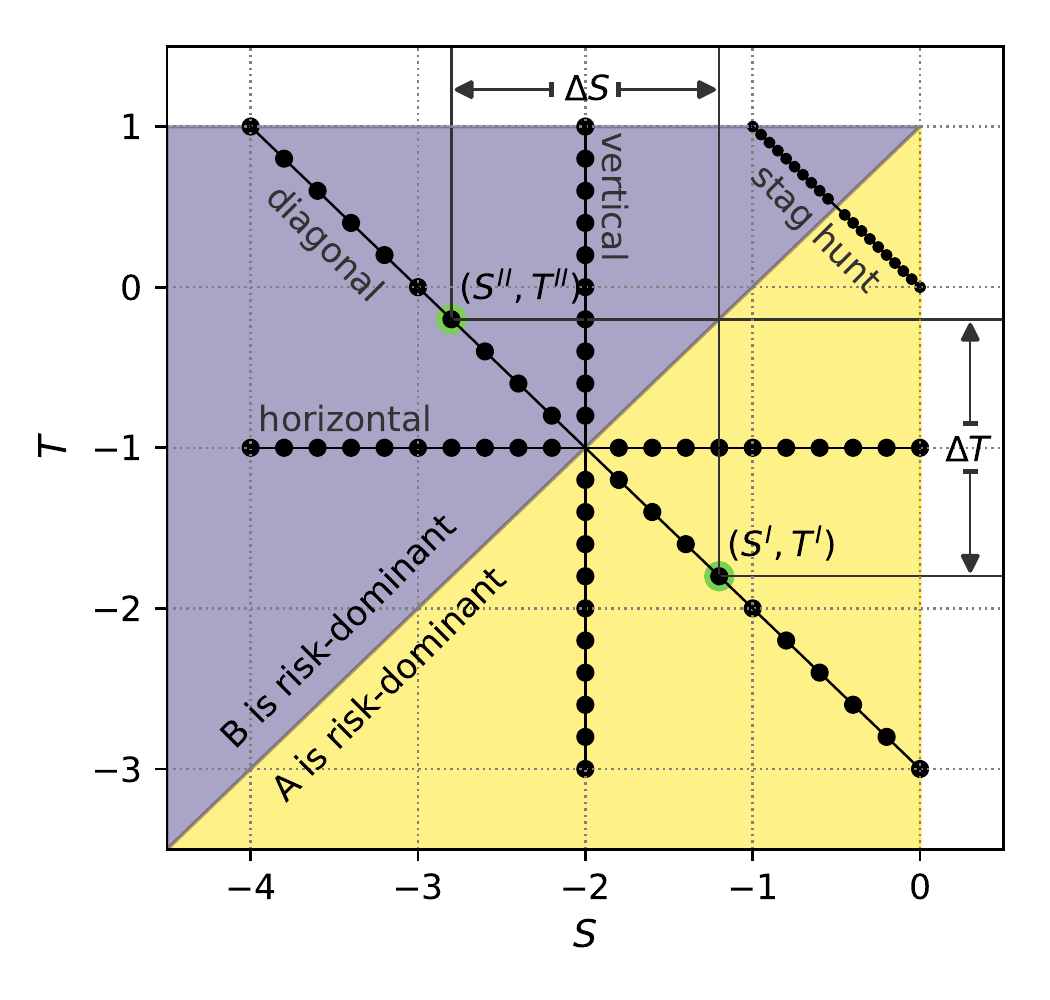}
    \caption{Diagram of the $S$-$T$ parameter space showing parametrisation
of the layers. Each circle on the diagonal lines represents a game
played on one of the layers. On layer I the strategy A is always
risk-dominant (yellow area),
and on layer II the strategy B is always risk-dominant (purpule area).
Risk-dominance changes at the line $T=S+1$. Examplary values
of $(S^I, T^I)$ and $(S^{II}, T^{II})$ are highlighted in green
with $\Delta S$ and $\Delta T$ illustrated.}
    \label{fig:diagram}
\end{figure}

\newpage

\restoregeometry
\newgeometry{tmargin=0.4in, bmargin=0.2in, lmargin=1.5cm, rmargin=1.5cm}

\newpage
\begin{figure}[th!]
\centerline{
    \includegraphics[scale=0.6]{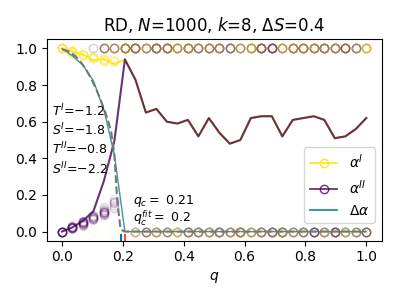}
    \includegraphics[scale=0.6]{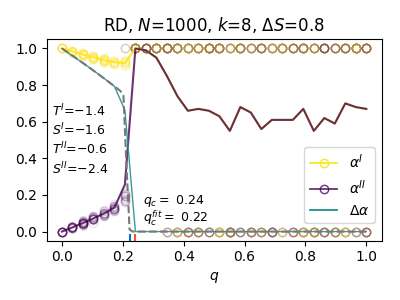}
    \includegraphics[scale=0.6]{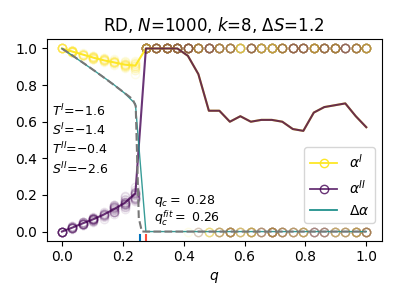}
    }
\centerline{
    \includegraphics[scale=0.6]{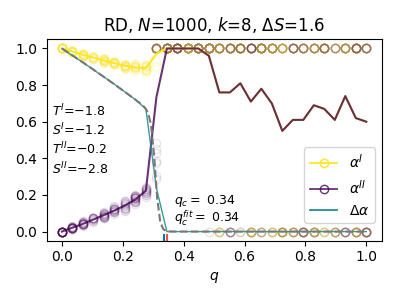}
    \includegraphics[scale=0.6]{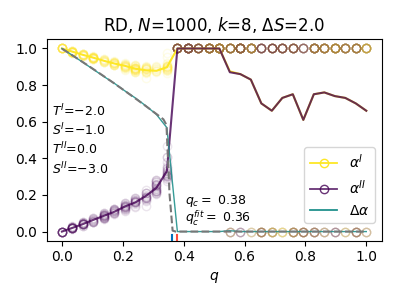}
    \includegraphics[scale=0.6]{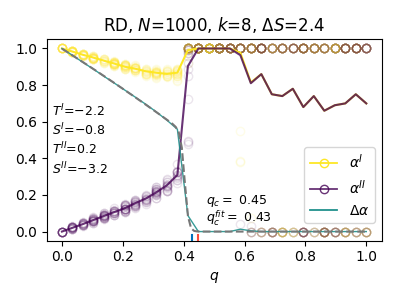}
    }
\centerline{
    \includegraphics[scale=0.6]{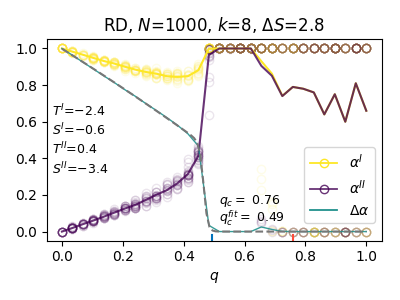}
    \includegraphics[scale=0.6]{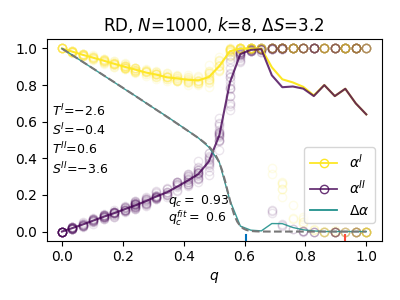}
    \includegraphics[scale=0.6]{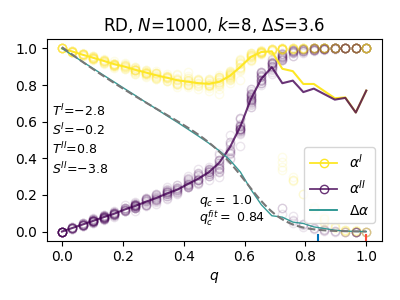}
    }
\centerline{
    \includegraphics[scale=0.6]{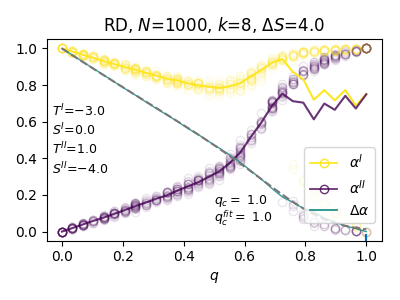}
    \includegraphics[scale=0.6]{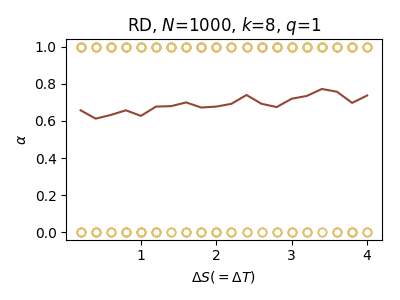}
       \includegraphics[scale=0.6]{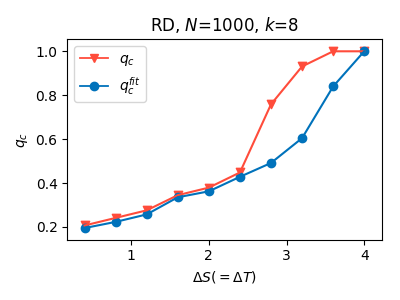}
 }
\centerline{
    \includegraphics[scale=0.7]{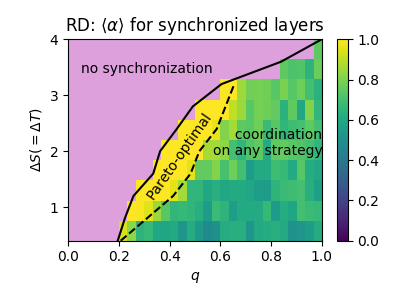}
    }
    \vspace{-0.5cm}
\caption{\footnotesize \textbf{RD in the diagonal case}
(i) Coordination rates on layers $\alpha^{I}$, $\alpha^{II}$,
and $\Delta \alpha$ vs node overlap $q$ for all
values of $\Delta S$.
(ii) Coordination rate $\alpha=\alpha^I=\alpha^{II}$
vs gap size $\Delta S$ for full node overlap $q=1$.
(iii) Critical value of $q_c$ and $q_c^{fit}$ vs gap size $\Delta S$.
(iv) Phase diagram of coordination rate $\alpha=\alpha^I=\alpha^{II}$
in the $q$-$\Delta S$ space for synchronised layers.
Each layer has $N=1000$ nodes with an intra-layer degree $k=8$.
Averaged over (at least) 100 realisations.
}
\end{figure}

\newpage
\begin{figure}[th!]
\centerline{
    \includegraphics[scale=0.6]{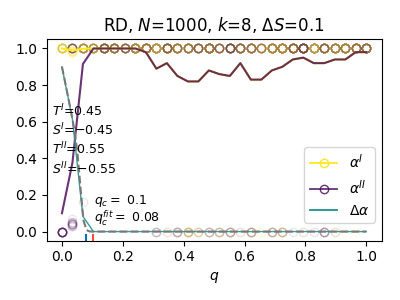}
    \includegraphics[scale=0.6]{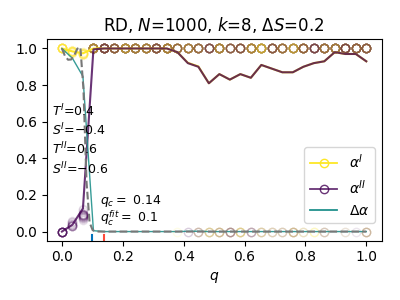}
    \includegraphics[scale=0.6]{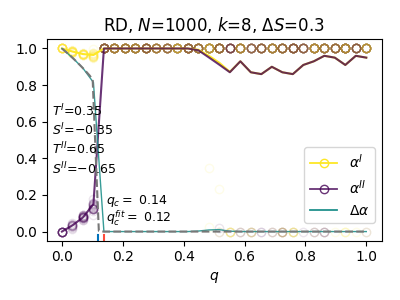}
    }
\centerline{
    \includegraphics[scale=0.6]{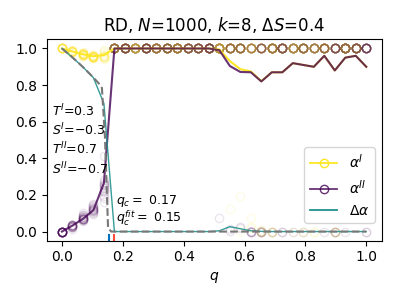}
    \includegraphics[scale=0.6]{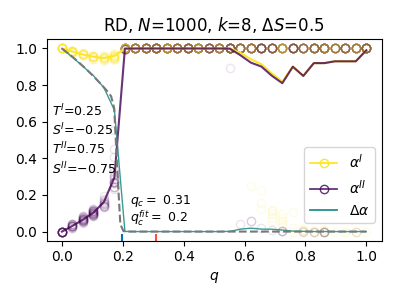}
    \includegraphics[scale=0.6]{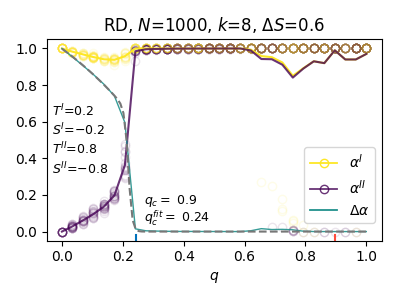}
    }
\centerline{
    \includegraphics[scale=0.6]{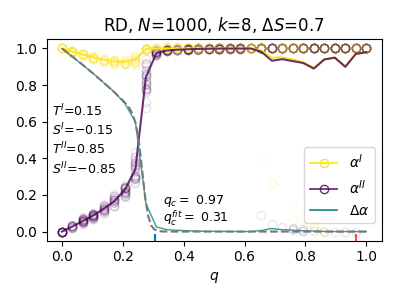}
    \includegraphics[scale=0.6]{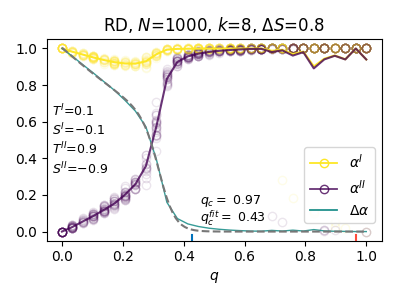}
    \includegraphics[scale=0.6]{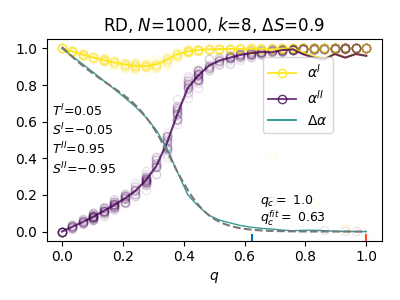}
    }
\centerline{
    \includegraphics[scale=0.6]{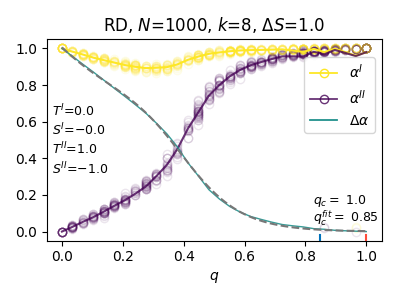}
    \includegraphics[scale=0.6]{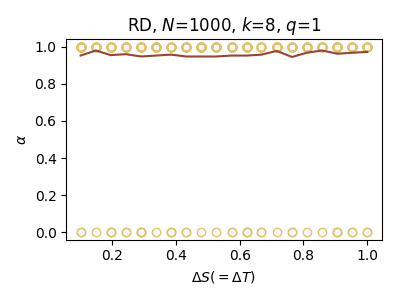}
       \includegraphics[scale=0.6]{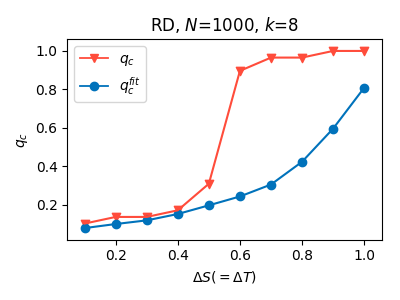}
 }
\centerline{
    \includegraphics[scale=0.7]{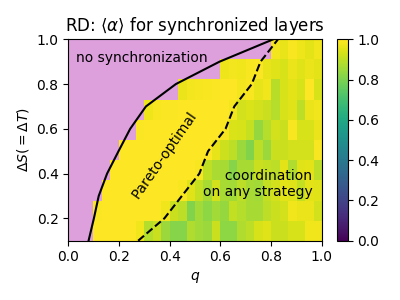}
    }
    \vspace{-0.5cm}
\caption{\footnotesize \textbf{RD in the stag hunt case}
(i) Coordination rates on layers $\alpha^{I}$, $\alpha^{II}$,
and $\Delta \alpha$ vs node overlap $q$ for all
values of $\Delta S$.
(ii) Coordination rate $\alpha=\alpha^I=\alpha^{II}$
vs gap size $\Delta S$ for full node overlap $q=1$.
(iii) Critical value of $q_c$ and $q_c^{fit}$ vs gap size $\Delta S$.
(iv) Phase diagram of coordination rate $\alpha=\alpha^I=\alpha^{II}$
in the $q$-$\Delta S$ space for synchronised layers.
Each layer has $N=1000$ nodes with an intra-layer degree $k=8$.
Averaged over (at least) 100 realisations.
}\end{figure}

\newpage
\begin{figure}[th!]
\centerline{
    \includegraphics[scale=0.6]{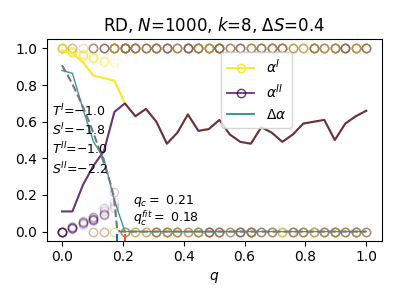}
    \includegraphics[scale=0.6]{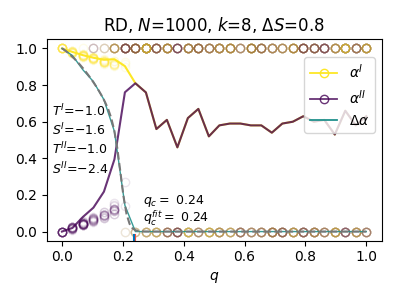}
    \includegraphics[scale=0.6]{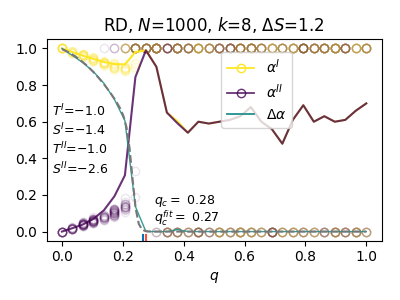}
}
\centerline{
    \includegraphics[scale=0.6]{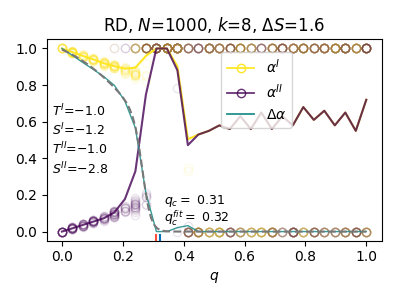}
    \includegraphics[scale=0.6]{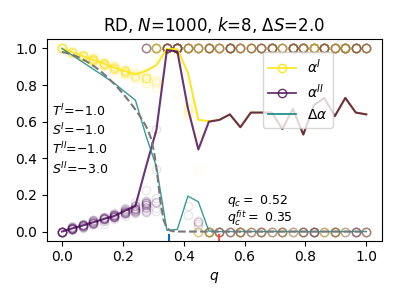}
    \includegraphics[scale=0.6]{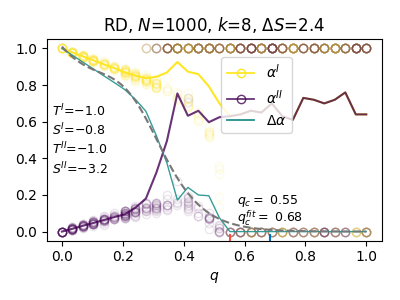}
    }
\centerline{
    \includegraphics[scale=0.6]{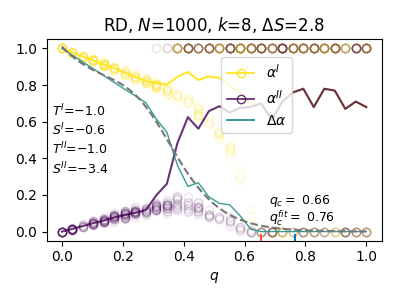}
    \includegraphics[scale=0.6]{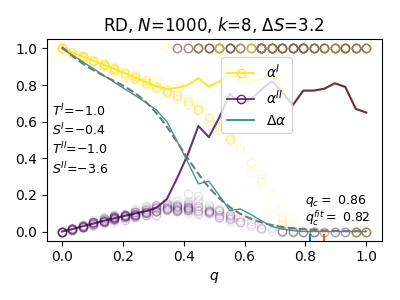}
    \includegraphics[scale=0.6]{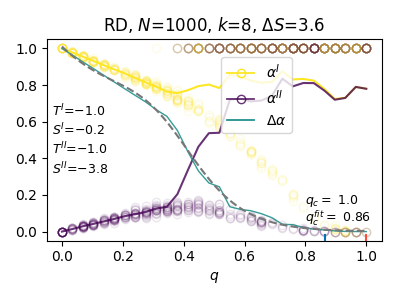}
    }
\centerline{
    \includegraphics[scale=0.6]{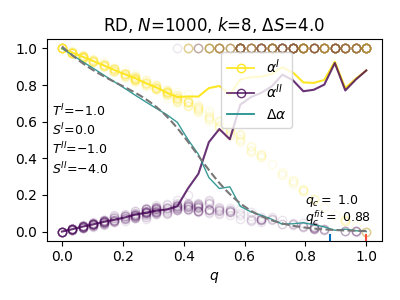}
    \includegraphics[scale=0.6]{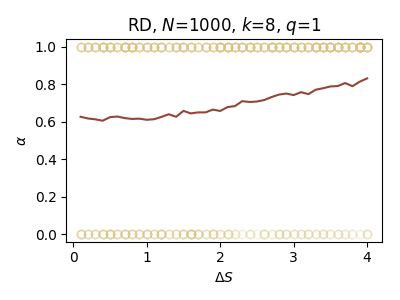}
    \includegraphics[scale=0.6]{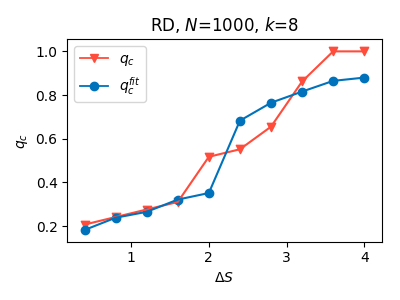}
}
\centerline{
    \includegraphics[scale=0.7]{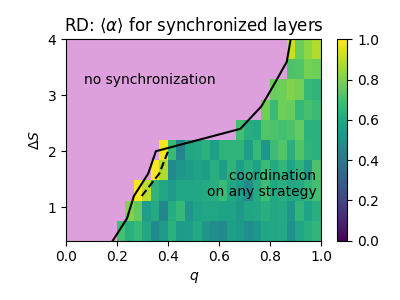}
}
    \vspace{-0.5cm}
\caption{\footnotesize \textbf{RD in the horizontal case}
(i) Coordination rates on layers $\alpha^{I}$, $\alpha^{II}$,
and $\Delta \alpha$ vs node overlap $q$ for all
values of $\Delta S$.
(ii) Coordination rate $\alpha=\alpha^I=\alpha^{II}$
vs gap size $\Delta S$ for full node overlap $q=1$.
(iii) Critical value of $q_c$ and $q_c^{fit}$ vs gap size $\Delta S$.
(iv) Phase diagram of coordination rate $\alpha=\alpha^I=\alpha^{II}$
in the $q$-$\Delta S$ space for synchronised layers.
Each layer has $N=1000$ nodes with an intra-layer degree $k=8$.
Averaged over (at least) 100 realisations.
}\end{figure}

\newpage
\begin{figure}[th!]
\centerline{
    \includegraphics[scale=0.6]{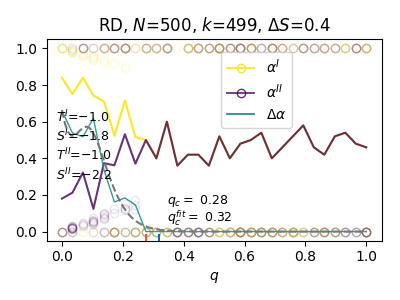}
    \includegraphics[scale=0.6]{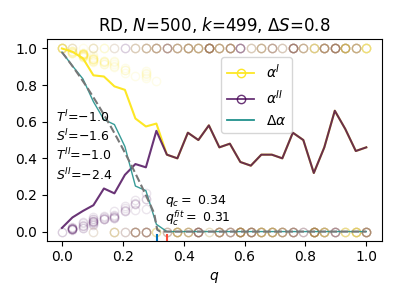}
    \includegraphics[scale=0.6]{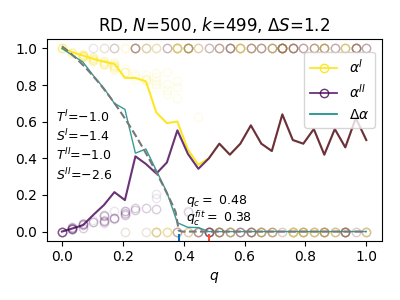}
    }
\centerline{
    \includegraphics[scale=0.6]{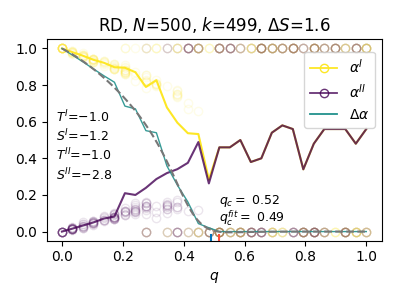}
    \includegraphics[scale=0.6]{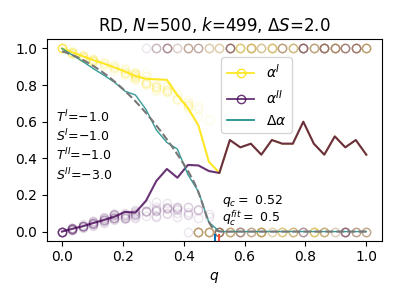}
    \includegraphics[scale=0.6]{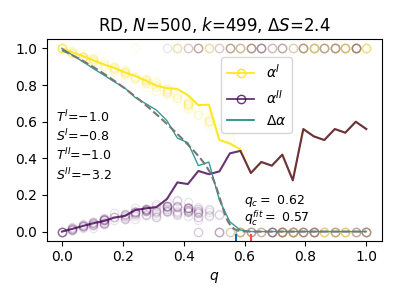}
    }
\centerline{
    \includegraphics[scale=0.6]{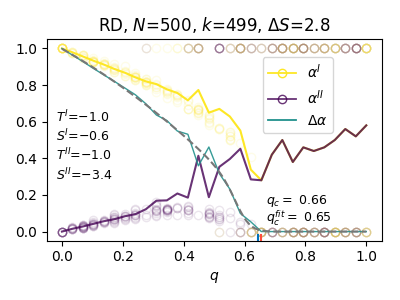}
    \includegraphics[scale=0.6]{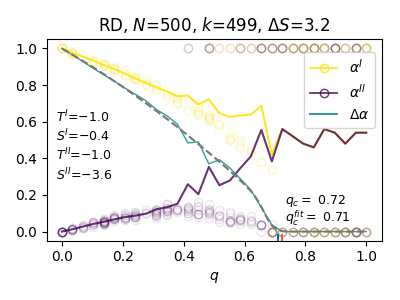}
    \includegraphics[scale=0.6]{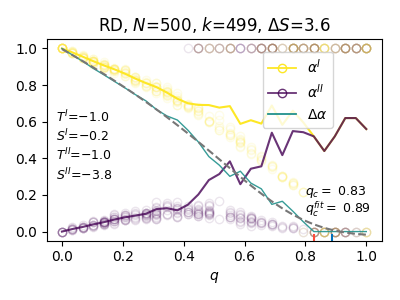}
    }
\centerline{
    \includegraphics[scale=0.6]{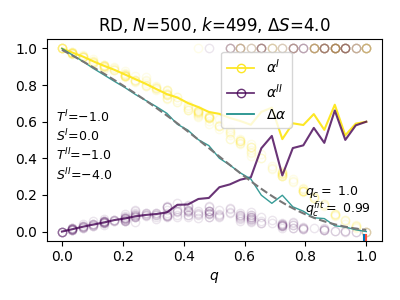}
    \includegraphics[scale=0.6]{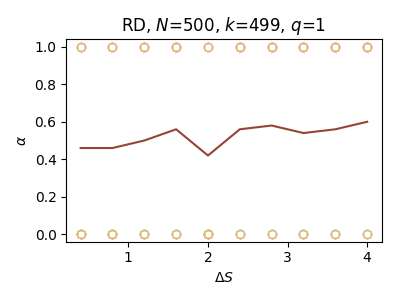}
        \includegraphics[scale=0.6]{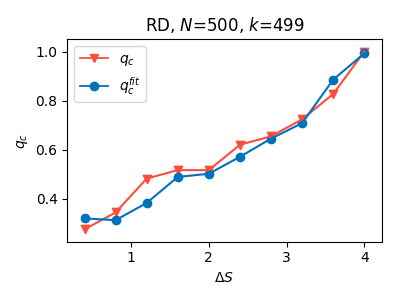}
    }
\centerline{
    \includegraphics[scale=0.7]{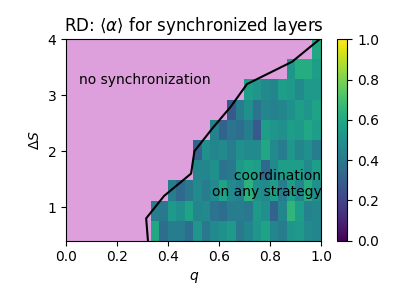}
    }
    \vspace{-0.5cm}
\caption{\footnotesize \textbf{RD in the horizontal case}
(i) Coordination rates on layers $\alpha^{I}$, $\alpha^{II}$,
and $\Delta \alpha$ vs node overlap $q$ for all
values of $\Delta S$.
(ii) Coordination rate $\alpha=\alpha^I=\alpha^{II}$
vs gap size $\Delta S$ for full node overlap $q=1$.
(iii) Critical value of $q_c$ and $q_c^{fit}$ vs gap size $\Delta S$.
(iv) Phase diagram of coordination rate $\alpha=\alpha^I=\alpha^{II}$
in the $q$-$\Delta S$ space for synchronised layers.
Each layer has $N=500$ nodes and forms a complete graph.
Averaged over (at least) 50 realisations.
}\end{figure}

\newpage
\begin{figure}[th!]
\centerline{
    \includegraphics[scale=0.6]{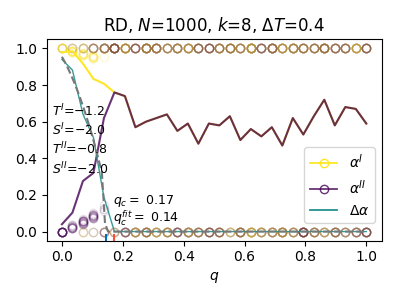}
    \includegraphics[scale=0.6]{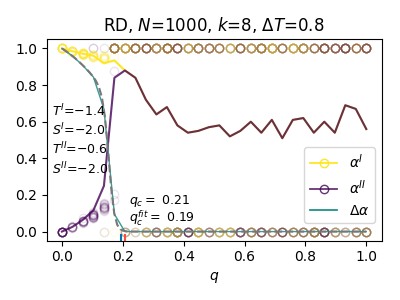}
    \includegraphics[scale=0.6]{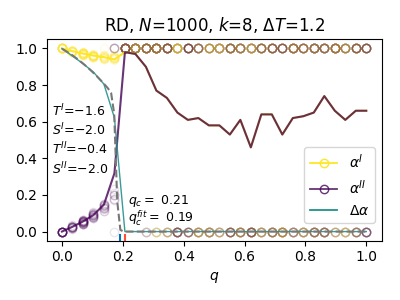}
    }
\centerline{
    \includegraphics[scale=0.6]{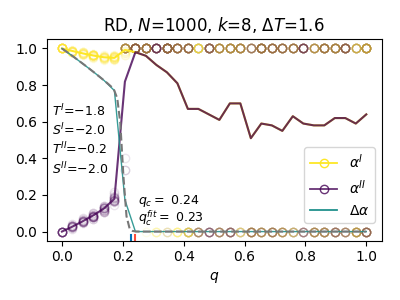}
    \includegraphics[scale=0.6]{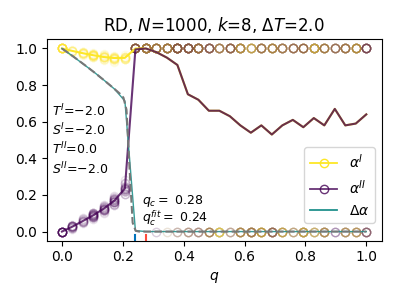}
    \includegraphics[scale=0.6]{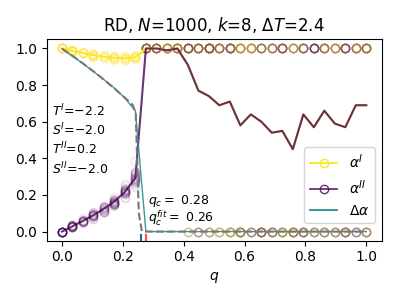}
    }
\centerline{
    \includegraphics[scale=0.6]{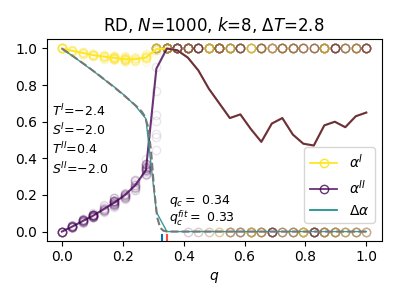}
    \includegraphics[scale=0.6]{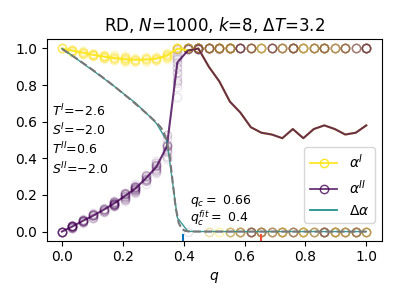}
    \includegraphics[scale=0.6]{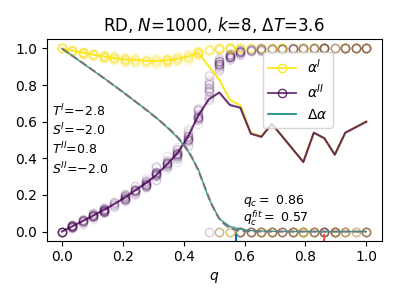}
    }
\centerline{
    \includegraphics[scale=0.6]{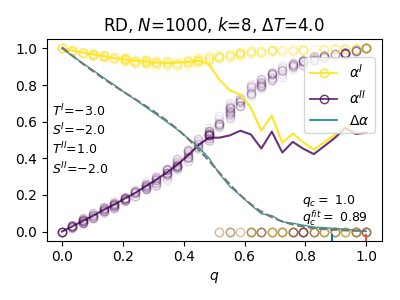}
    \includegraphics[scale=0.6]{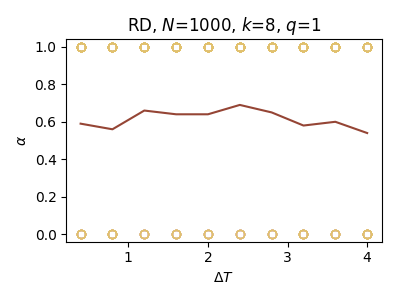}
    \includegraphics[scale=0.6]{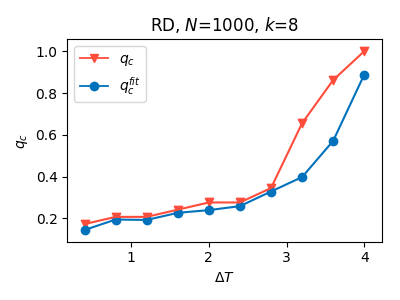}
    }
\centerline{
    \includegraphics[scale=0.7]{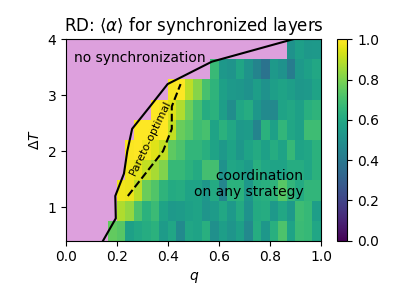}
    }
    \vspace{-0.5cm}
\caption{\footnotesize \textbf{RD in the vertical case}
(i) Coordination rates on layers $\alpha^{I}$, $\alpha^{II}$,
and $\Delta \alpha$ vs node overlap $q$ for all
values of $\Delta T$.
(ii) Coordination rate $\alpha=\alpha^I=\alpha^{II}$
vs gap size $\Delta T$ for full node overlap $q=1$.
(iii) Critical value of $q_c$ and $q_c^{fit}$ vs gap size $\Delta T$.
(iv) Phase diagram of coordination rate $\alpha=\alpha^I=\alpha^{II}$
in the $q$-$\Delta T$ space for synchronised layers.
Each layer has $N=1000$ nodes with an intra-layer degree $k=8$.
Averaged over (at least) 100 realisations.
}\end{figure}


\newpage
\begin{figure}[th!]
\centerline{
    \includegraphics[scale=0.6]{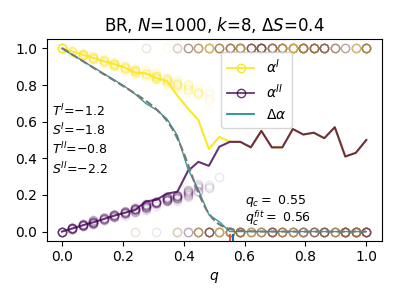}
    \includegraphics[scale=0.6]{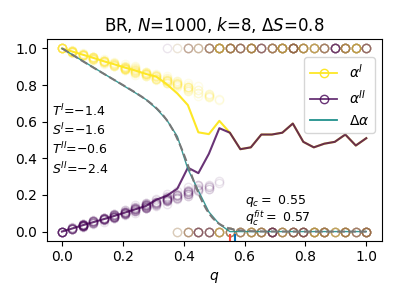}
    \includegraphics[scale=0.6]{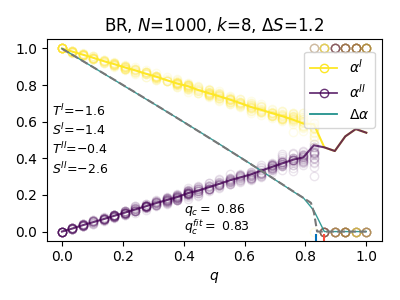}
    }
\centerline{
    \includegraphics[scale=0.6]{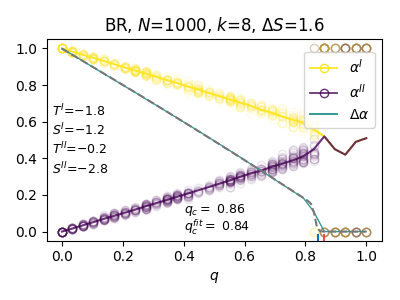}
    \includegraphics[scale=0.6]{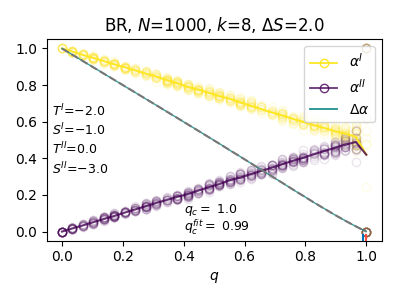}
    \includegraphics[scale=0.6]{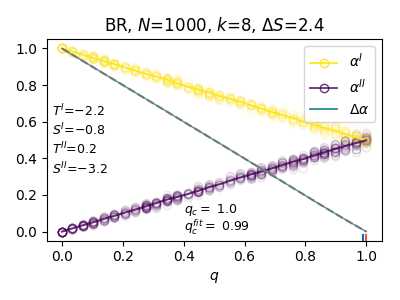}
    }
\centerline{
    \includegraphics[scale=0.6]{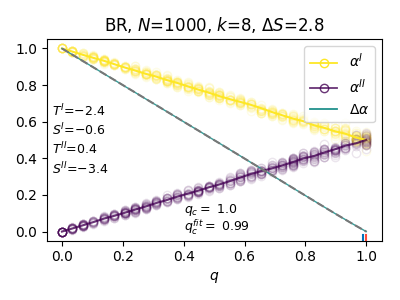}
    \includegraphics[scale=0.6]{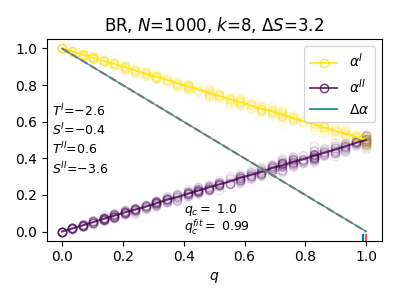}
    \includegraphics[scale=0.6]{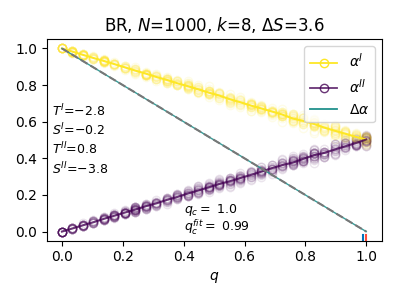}
    }
\centerline{
    \includegraphics[scale=0.6]{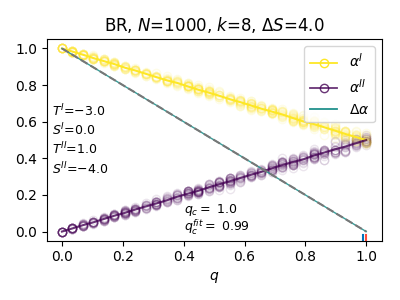}
    \includegraphics[scale=0.6]{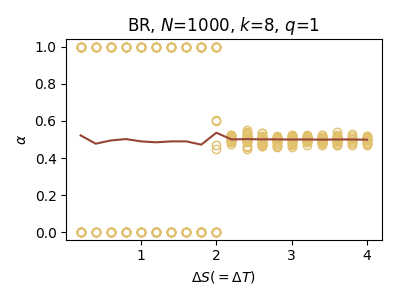}
       \includegraphics[scale=0.6]{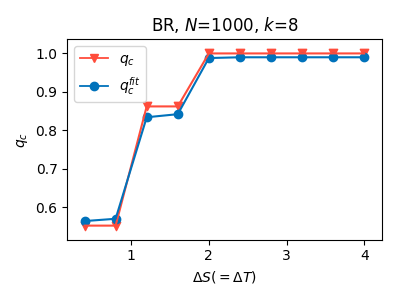}
 }
\centerline{
    \includegraphics[scale=0.7]{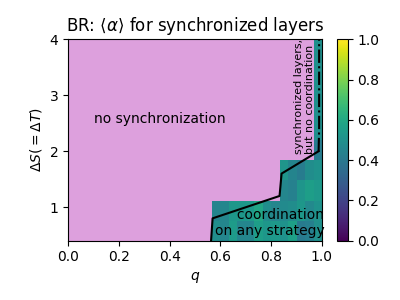}
    }
    \vspace{-0.5cm}
\caption{\footnotesize \textbf{BR in the diagonal case}
(i) Coordination rates on layers $\alpha^{I}$, $\alpha^{II}$,
and $\Delta \alpha$ vs node overlap $q$ for all
values of $\Delta S$.
(ii) Coordination rate $\alpha=\alpha^I=\alpha^{II}$
vs gap size $\Delta S$ for full node overlap $q=1$.
(iii) Critical value of $q_c$ and $q_c^{fit}$ vs gap size $\Delta S$.
(iv) Phase diagram of coordination rate $\alpha=\alpha^I=\alpha^{II}$
in the $q$-$\Delta S$ space for synchronised layers.
Each layer has $N=1000$ nodes with an intra-layer degree $k=8$.
Averaged over (at least) 100 realisations.
}
\end{figure}

\newpage
\begin{figure}[th!]
\centerline{
    \includegraphics[scale=0.6]{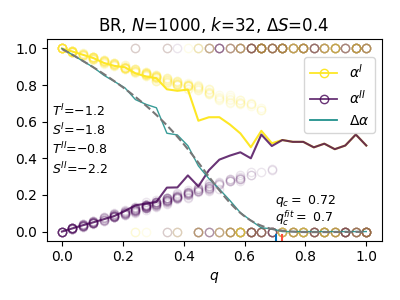}
    \includegraphics[scale=0.6]{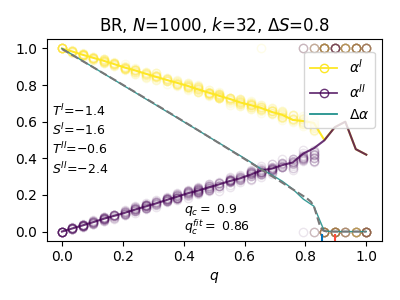}
    \includegraphics[scale=0.6]{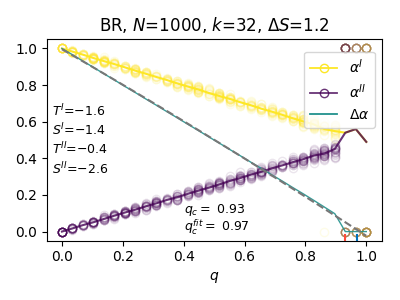}
    }
\centerline{
    \includegraphics[scale=0.6]{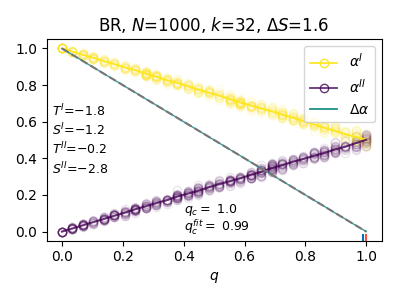}
    \includegraphics[scale=0.6]{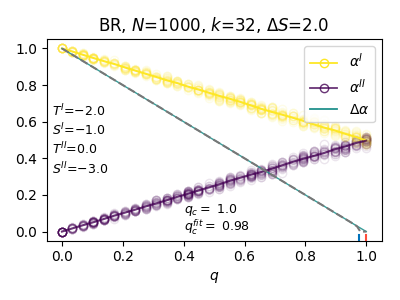}
    \includegraphics[scale=0.6]{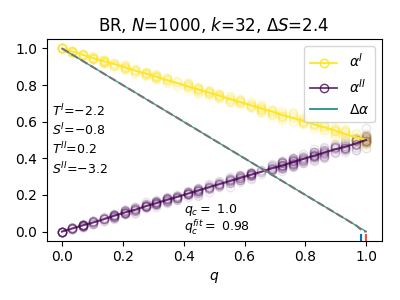}
    }
\centerline{
    \includegraphics[scale=0.6]{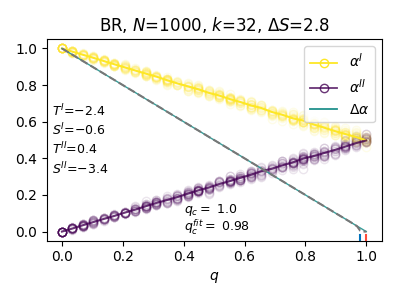}
    \includegraphics[scale=0.6]{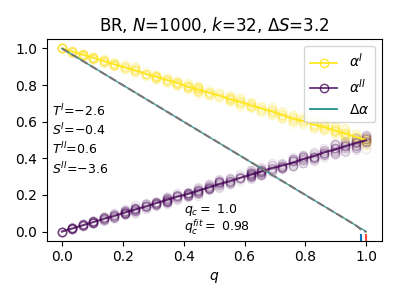}
    \includegraphics[scale=0.6]{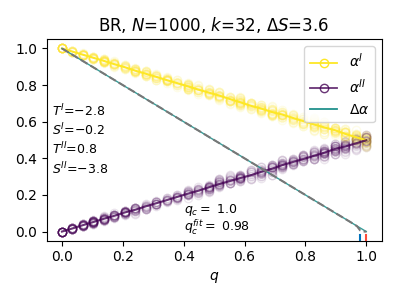}
    }
\centerline{
    \includegraphics[scale=0.6]{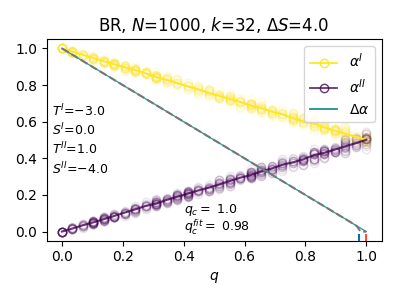}
    \includegraphics[scale=0.6]{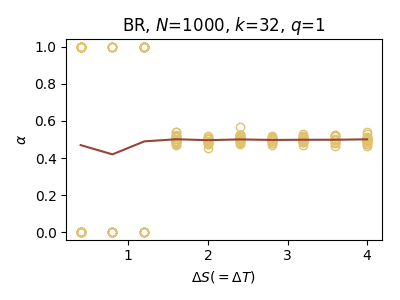}
      \includegraphics[scale=0.6]{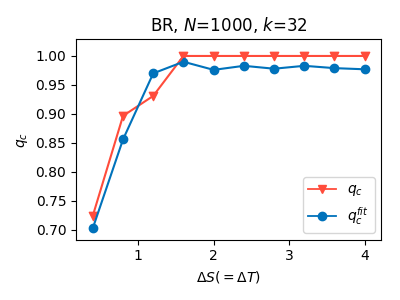}
  }
\centerline{
    \includegraphics[scale=0.7]{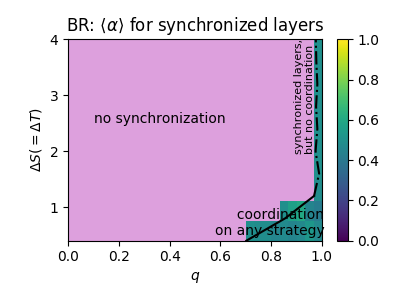}
    }
    \vspace{-0.5cm}
\caption{\footnotesize \textbf{BR in the diagonal case}
(i) Coordination rates on layers $\alpha^{I}$, $\alpha^{II}$,
and $\Delta \alpha$ vs node overlap $q$ for all
values of $\Delta S$.
(ii) Coordination rate $\alpha=\alpha^I=\alpha^{II}$
vs gap size $\Delta S$ for full node overlap $q=1$.
(iii) Critical value of $q_c$ and $q_c^{fit}$ vs gap size $\Delta S$.
(iv) Phase diagram of coordination rate $\alpha=\alpha^I=\alpha^{II}$
in the $q$-$\Delta S$ space for synchronised layers.
Each layer has $N=1000$ nodes with an intra-layer degree $k=32$.
Averaged over (at least) 100 realisations.
}\end{figure}

\newpage
\begin{figure}[th!]
\centerline{
    \includegraphics[scale=0.6]{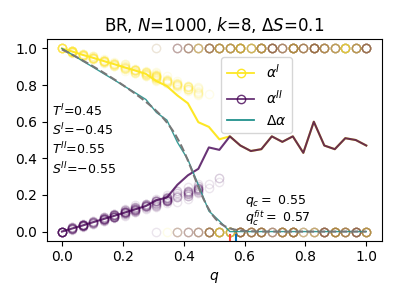}
    \includegraphics[scale=0.6]{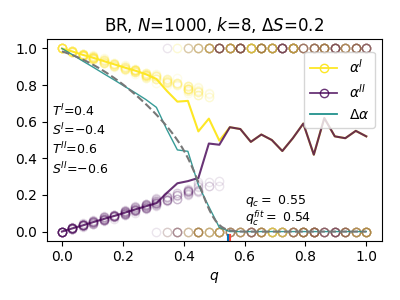}
    \includegraphics[scale=0.6]{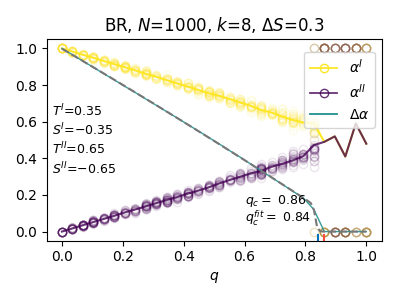}
    }
\centerline{
    \includegraphics[scale=0.6]{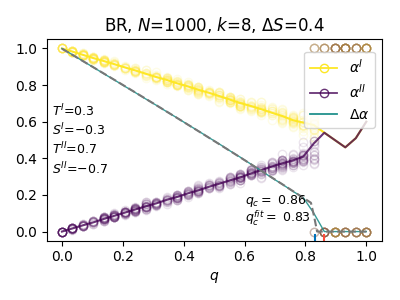}
    \includegraphics[scale=0.6]{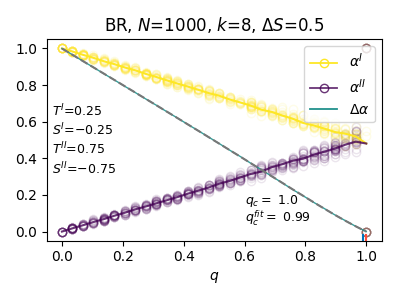}
    \includegraphics[scale=0.6]{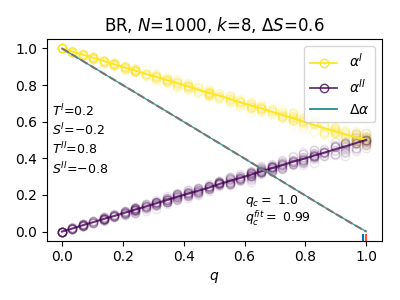}
    }
\centerline{
    \includegraphics[scale=0.6]{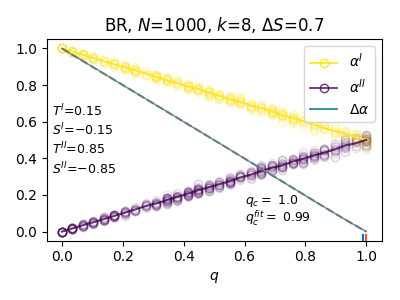}
    \includegraphics[scale=0.6]{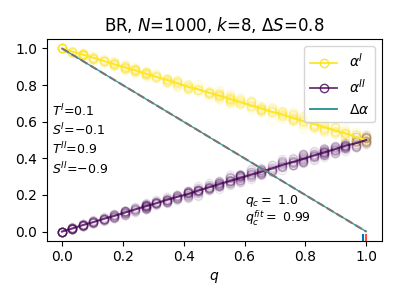}
    \includegraphics[scale=0.6]{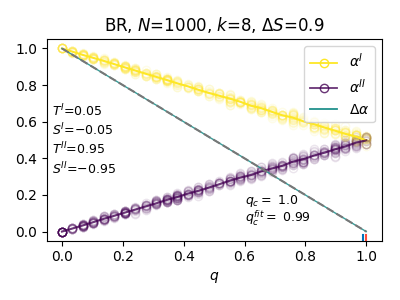}
    }
\centerline{
    \includegraphics[scale=0.6]{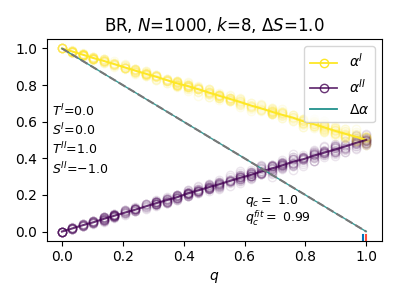}
    \includegraphics[scale=0.6]{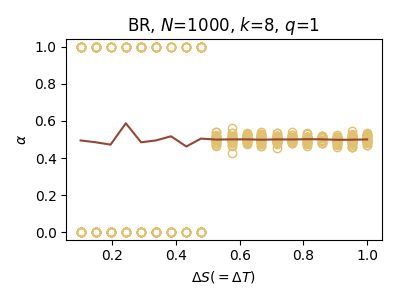}
       \includegraphics[scale=0.6]{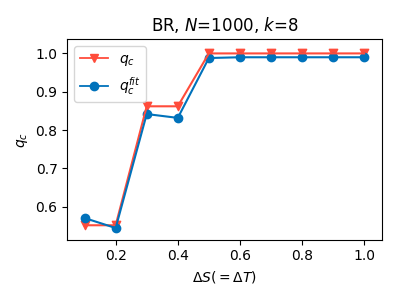}
 }
\centerline{
    \includegraphics[scale=0.7]{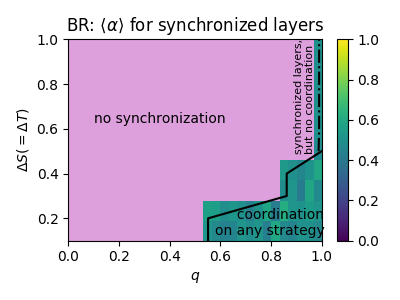}
    }
    \vspace{-0.5cm}
\caption{\footnotesize \textbf{BR in the stag hunt case}
(i) Coordination rates on layers $\alpha^{I}$, $\alpha^{II}$,
and $\Delta \alpha$ vs node overlap $q$ for all
values of $\Delta S$.
(ii) Coordination rate $\alpha=\alpha^I=\alpha^{II}$
vs gap size $\Delta S$ for full node overlap $q=1$.
(iii) Critical value of $q_c$ and $q_c^{fit}$ vs gap size $\Delta S$.
(iv) Phase diagram of coordination rate $\alpha=\alpha^I=\alpha^{II}$
in the $q$-$\Delta S$ space for synchronised layers.
Each layer has $N=1000$ nodes with an intra-layer degree $k=8$.
Averaged over (at least) 100 realisations.
}\end{figure}

\newpage
\begin{figure}[th!]
\centerline{
    \includegraphics[scale=0.6]{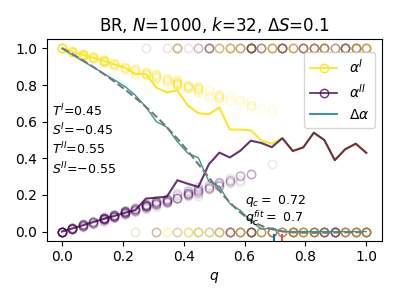}
    \includegraphics[scale=0.6]{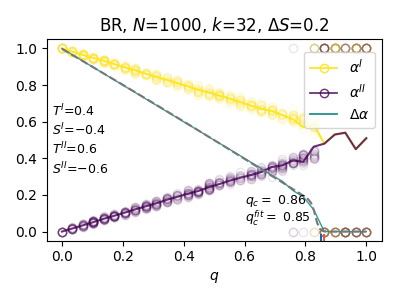}
    \includegraphics[scale=0.6]{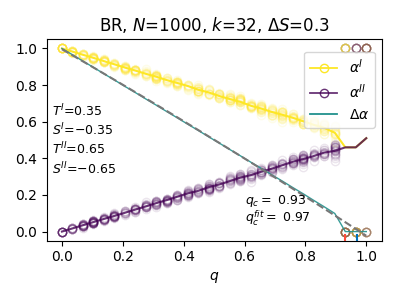}
    }
\centerline{
    \includegraphics[scale=0.6]{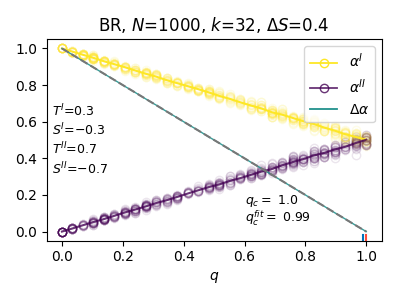}
    \includegraphics[scale=0.6]{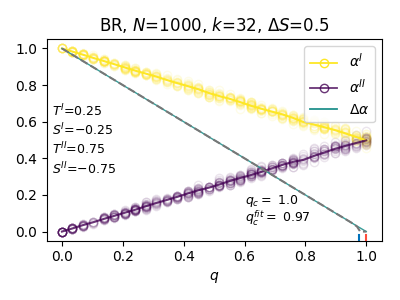}
    \includegraphics[scale=0.6]{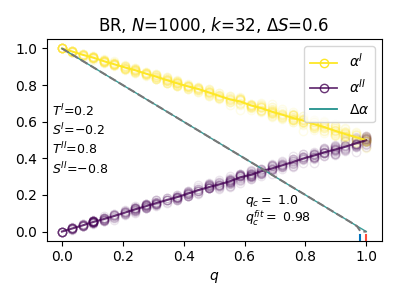}
    }
\centerline{
    \includegraphics[scale=0.6]{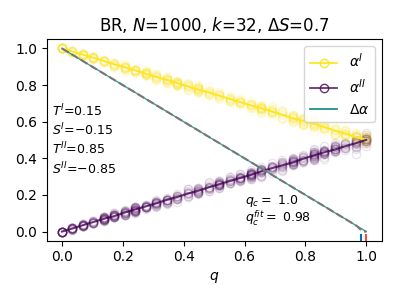}
    \includegraphics[scale=0.6]{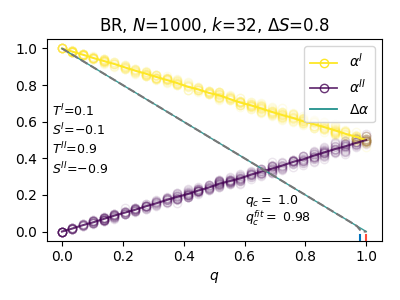}
    \includegraphics[scale=0.6]{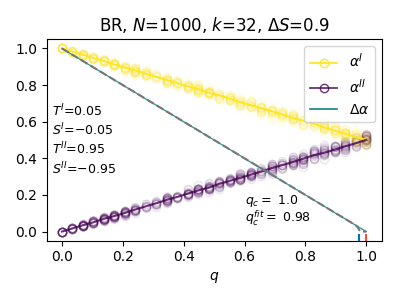}
    }
\centerline{
    \includegraphics[scale=0.6]{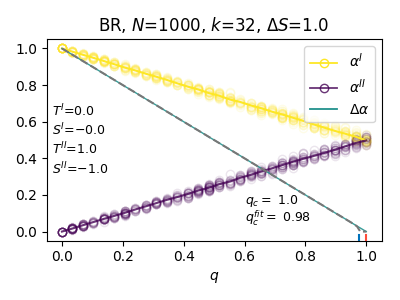}
    \includegraphics[scale=0.6]{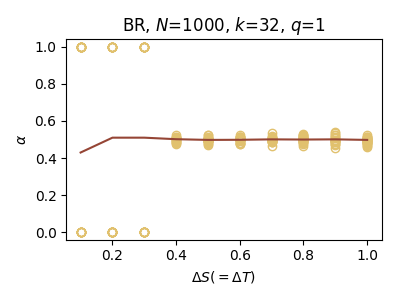}
       \includegraphics[scale=0.6]{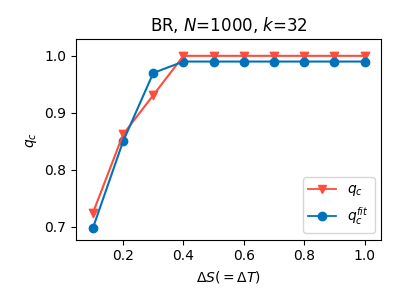}
 }
\centerline{
    \includegraphics[scale=0.7]{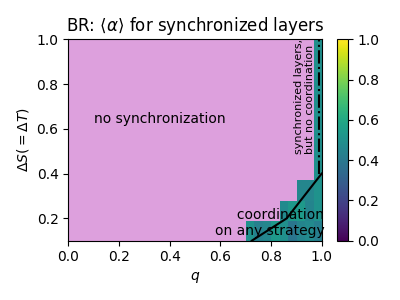}
    }
    \vspace{-0.5cm}
\caption{\footnotesize \textbf{BR in the stag hunt case}
(i) Coordination rates on layers $\alpha^{I}$, $\alpha^{II}$,
and $\Delta \alpha$ vs node overlap $q$ for all
values of $\Delta S$.
(ii) Coordination rate $\alpha=\alpha^I=\alpha^{II}$
vs gap size $\Delta S$ for full node overlap $q=1$.
(iii) Critical value of $q_c$ and $q_c^{fit}$ vs gap size $\Delta S$.
(iv) Phase diagram of coordination rate $\alpha=\alpha^I=\alpha^{II}$
in the $q$-$\Delta S$ space for synchronised layers.
Each layer has $N=1000$ nodes with an intra-layer degree $k=32$.
Averaged over (at least) 100 realisations.
}\end{figure}

\newpage
\begin{figure}[th!]
\centerline{
    \includegraphics[scale=0.6]{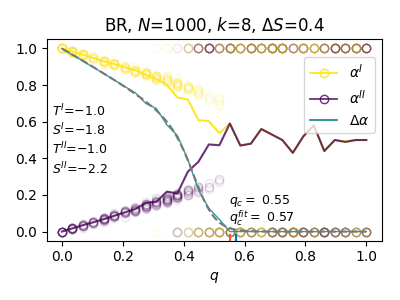}
    \includegraphics[scale=0.6]{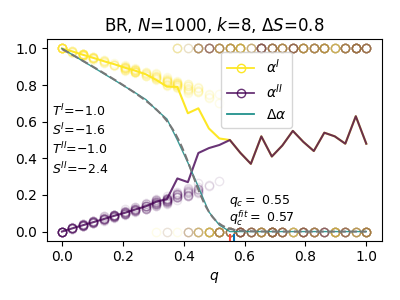}
    \includegraphics[scale=0.6]{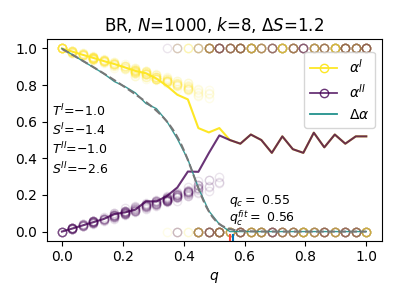}
    }
\centerline{
    \includegraphics[scale=0.6]{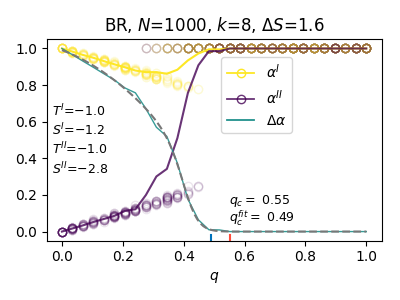}
    \includegraphics[scale=0.6]{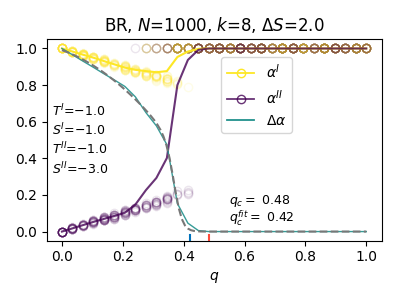}
    \includegraphics[scale=0.6]{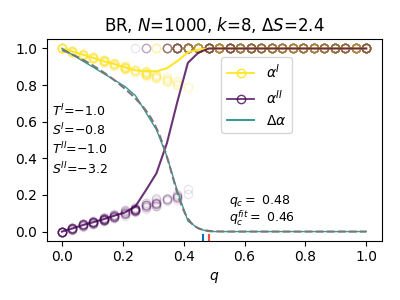}
    }
\centerline{
    \includegraphics[scale=0.6]{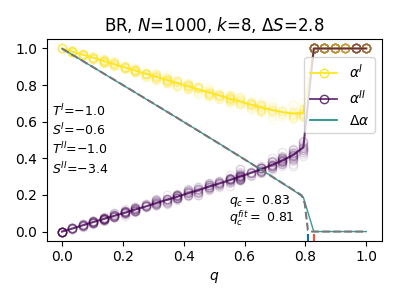}
    \includegraphics[scale=0.6]{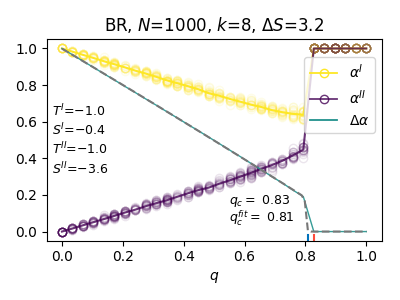}
    \includegraphics[scale=0.6]{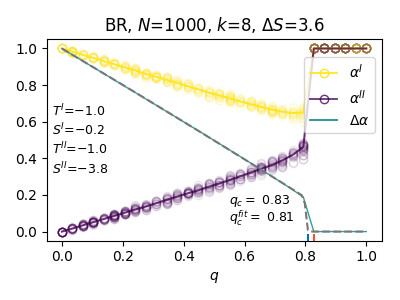}
    }
\centerline{
    \includegraphics[scale=0.6]{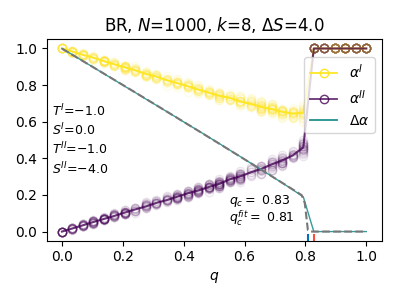}
    \includegraphics[scale=0.6]{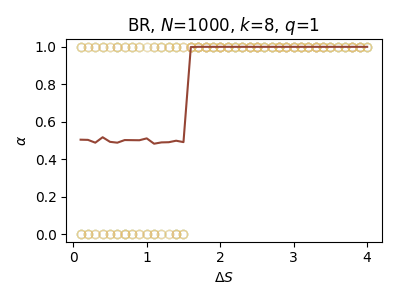}
       \includegraphics[scale=0.6]{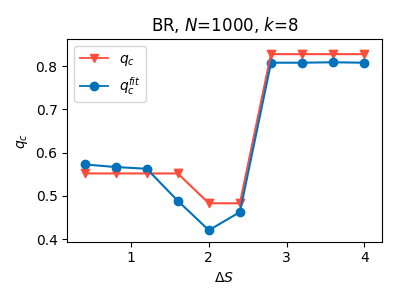}
 }
\centerline{
    \includegraphics[scale=0.7]{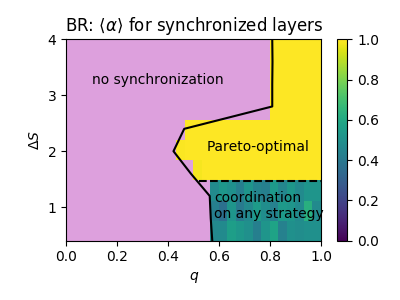}
    }
    \vspace{-0.5cm}
\caption{\footnotesize \textbf{BR in the horizontal case}
(i) Coordination rates on layers $\alpha^{I}$, $\alpha^{II}$,
and $\Delta \alpha$ vs node overlap $q$ for all
values of $\Delta S$.
(ii) Coordination rate $\alpha=\alpha^I=\alpha^{II}$
vs gap size $\Delta S$ for full node overlap $q=1$.
(iii) Critical value of $q_c$ and $q_c^{fit}$ vs gap size $\Delta S$.
(iv) Phase diagram of coordination rate $\alpha=\alpha^I=\alpha^{II}$
in the $q$-$\Delta S$ space for synchronised layers.
Each layer has $N=1000$ nodes with an intra-layer degree $k=8$.
Averaged over (at least) 100 realisations.
}\end{figure}

\newpage
\begin{figure}[th!]
\centerline{
    \includegraphics[scale=0.6]{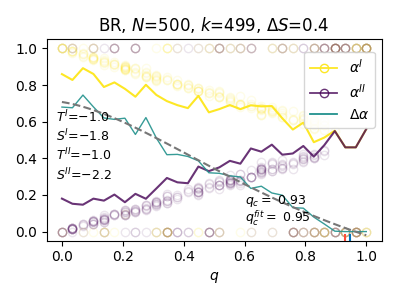}
    \includegraphics[scale=0.6]{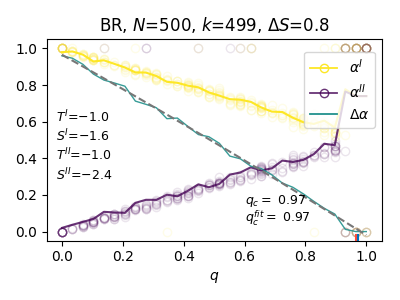}
    \includegraphics[scale=0.6]{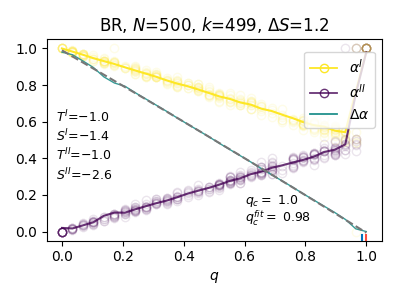}
    }
\centerline{
    \includegraphics[scale=0.6]{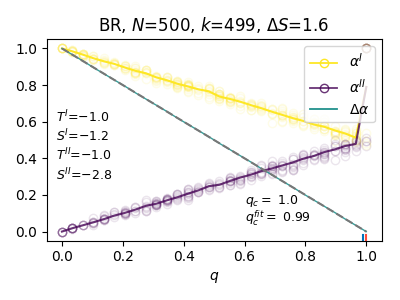}
    \includegraphics[scale=0.6]{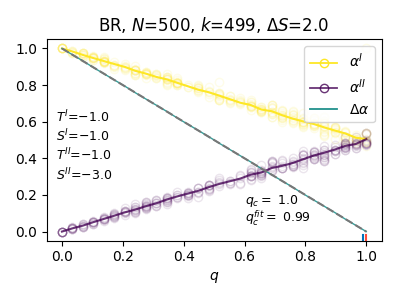}
    \includegraphics[scale=0.6]{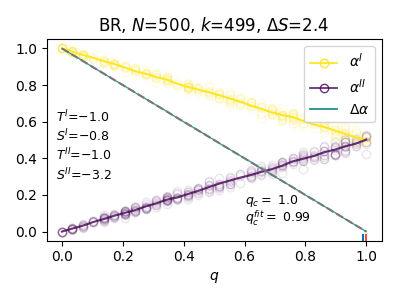}
    }
\centerline{
    \includegraphics[scale=0.6]{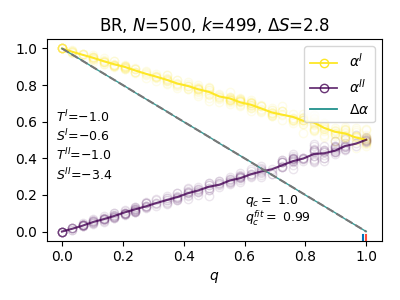}
    \includegraphics[scale=0.6]{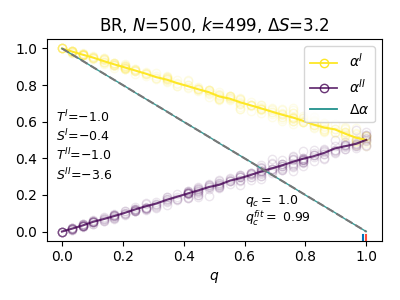}
    \includegraphics[scale=0.6]{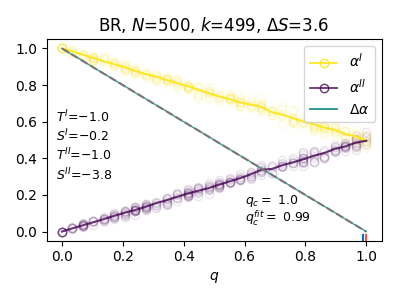}
    }
\centerline{
    \includegraphics[scale=0.6]{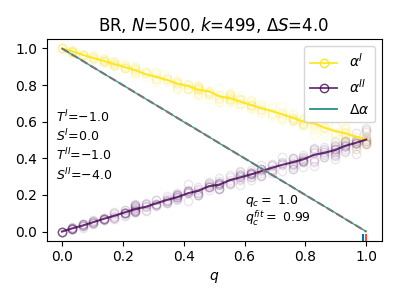}
    \includegraphics[scale=0.6]{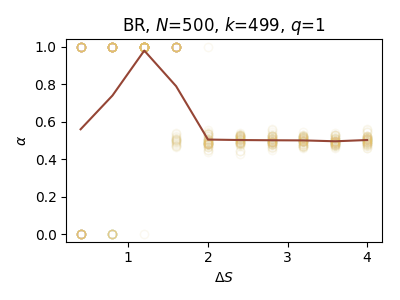}
        \includegraphics[scale=0.6]{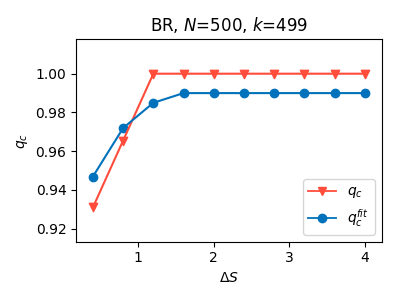}
    }
\centerline{
    \includegraphics[scale=0.7]{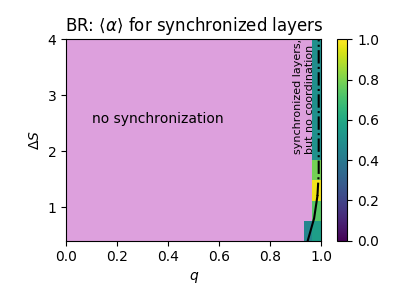}
    }
    \vspace{-0.5cm}
\caption{\footnotesize \textbf{BR in the horizontal case}
(i) Coordination rates on layers $\alpha^{I}$, $\alpha^{II}$,
and $\Delta \alpha$ vs node overlap $q$ for all
values of $\Delta S$.
(ii) Coordination rate $\alpha=\alpha^I=\alpha^{II}$
vs gap size $\Delta S$ for full node overlap $q=1$.
(iii) Critical value of $q_c$ and $q_c^{fit}$ vs gap size $\Delta S$.
(iv) Phase diagram of coordination rate $\alpha=\alpha^I=\alpha^{II}$
in the $q$-$\Delta S$ space for synchronised layers.
Each layer has $N=500$ nodes and forms a complete graph.
Averaged over (at least) 50 realisations.
}\end{figure}

\newpage
\begin{figure}[th!]
\centerline{
    \includegraphics[scale=0.6]{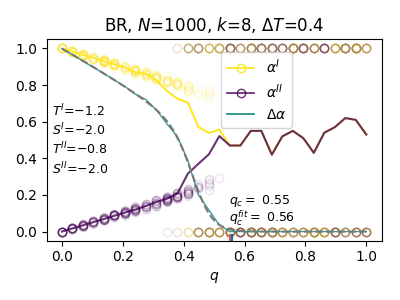}
    \includegraphics[scale=0.6]{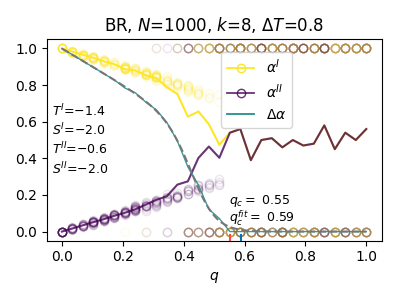}
    \includegraphics[scale=0.6]{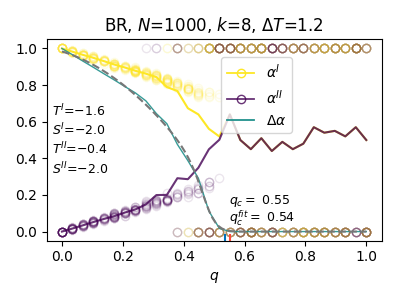}
    }
\centerline{
    \includegraphics[scale=0.6]{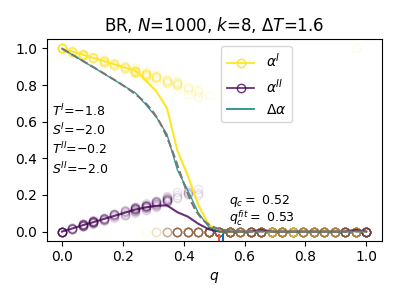}
    \includegraphics[scale=0.6]{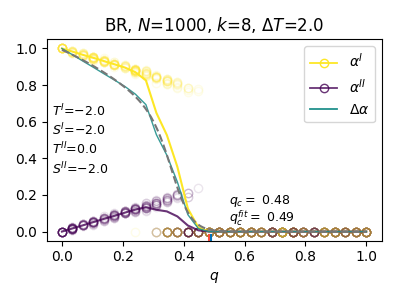}
    \includegraphics[scale=0.6]{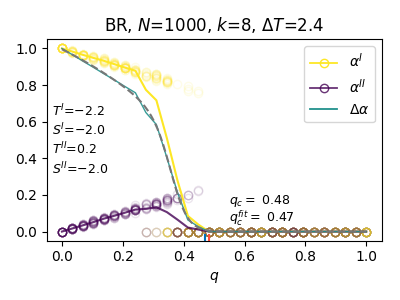}
    }
\centerline{
    \includegraphics[scale=0.6]{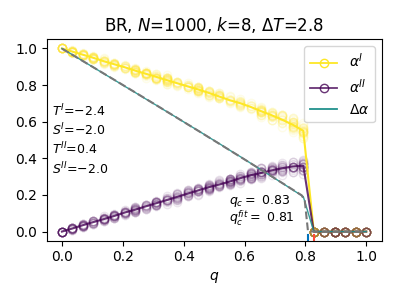}
    \includegraphics[scale=0.6]{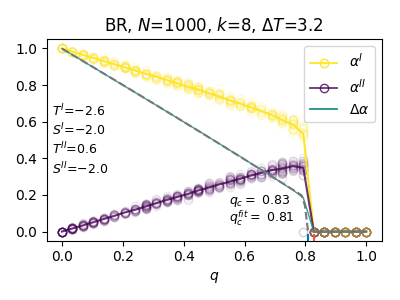}
    \includegraphics[scale=0.6]{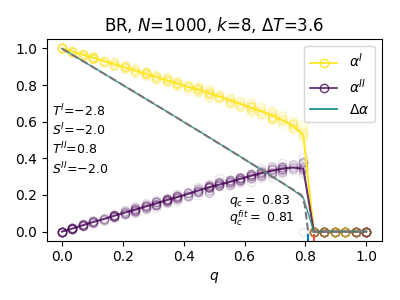}
    }
\centerline{
    \includegraphics[scale=0.6]{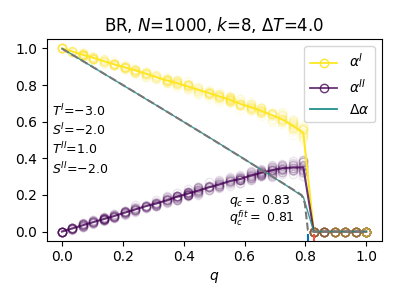}
    \includegraphics[scale=0.6]{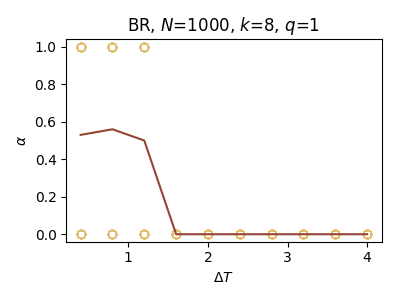}
    \includegraphics[scale=0.6]{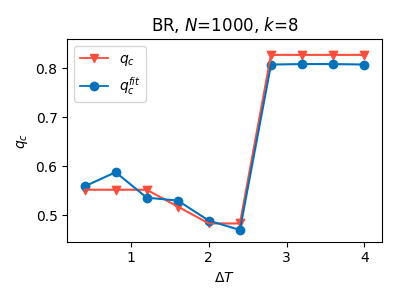}
    }
\centerline{
    \includegraphics[scale=0.7]{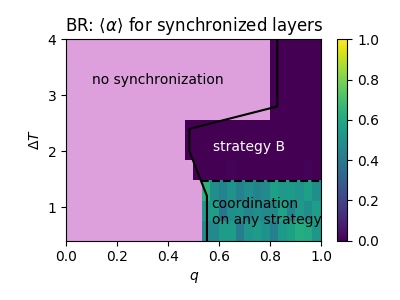}
    }
    \vspace{-0.5cm}
\caption{\footnotesize \textbf{BR in the vertical case}
(i) Coordination rates on layers $\alpha^{I}$, $\alpha^{II}$,
and $\Delta \alpha$ vs node overlap $q$ for all
values of $\Delta T$.
(ii) Coordination rate $\alpha=\alpha^I=\alpha^{II}$
vs gap size $\Delta T$ for full node overlap $q=1$.
(iii) Critical value of $q_c$ and $q_c^{fit}$ vs gap size $\Delta T$.
(iv) Phase diagram of coordination rate $\alpha=\alpha^I=\alpha^{II}$
in the $q$-$\Delta T$ space for synchronised layers.
Each layer has $N=1000$ nodes with an intra-layer degree $k=8$.
Averaged over (at least) 100 realisations.
}\end{figure}


\newpage
\begin{figure}[th!]
\centerline{
    \includegraphics[scale=0.6]{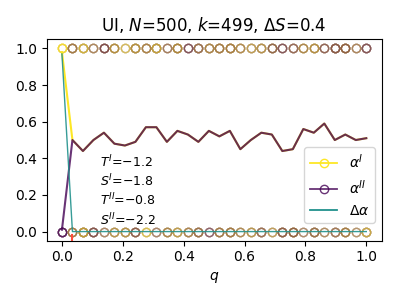}
    \includegraphics[scale=0.6]{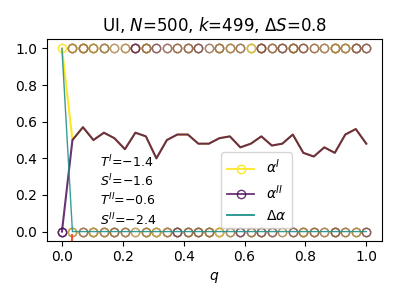}
    \includegraphics[scale=0.6]{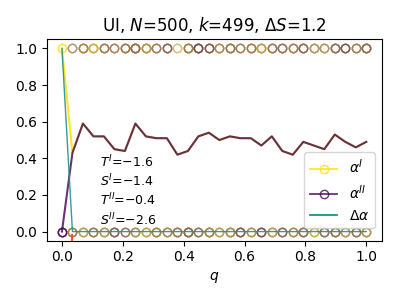}
    }
\centerline{
    \includegraphics[scale=0.6]{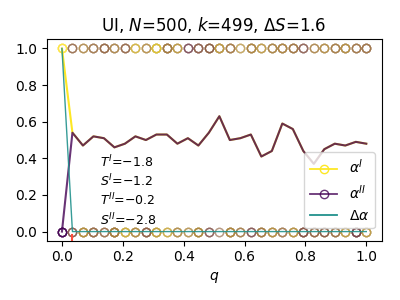}
    \includegraphics[scale=0.6]{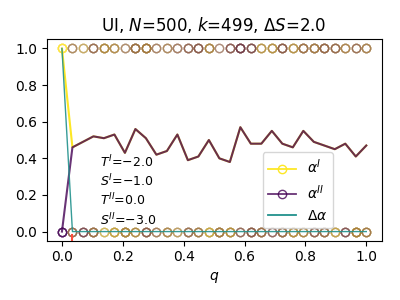}
    \includegraphics[scale=0.6]{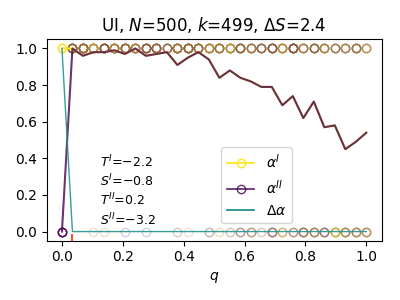}
    }
\centerline{
    \includegraphics[scale=0.6]{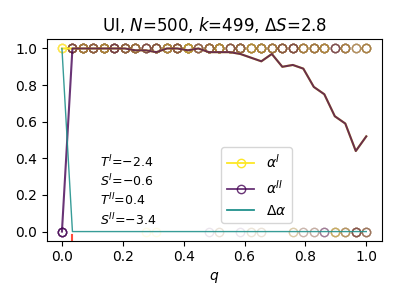}
    \includegraphics[scale=0.6]{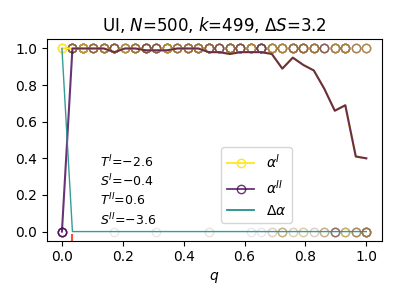}
    \includegraphics[scale=0.6]{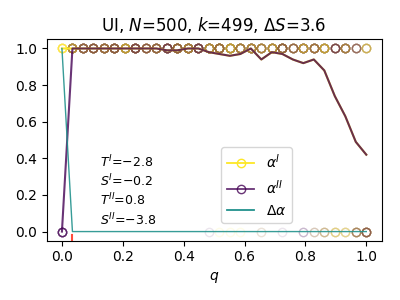}
    }
\centerline{
    \includegraphics[scale=0.6]{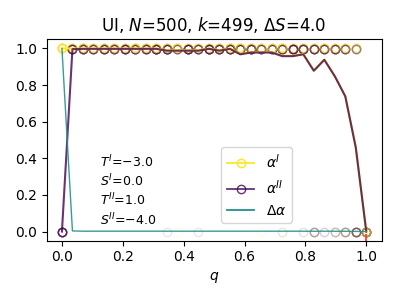}
    \includegraphics[scale=0.6]{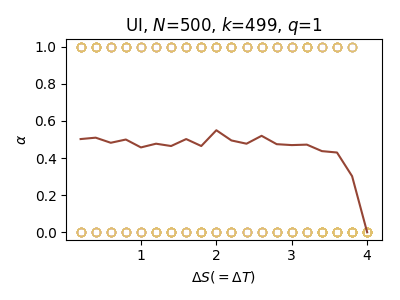}
    \includegraphics[scale=0.65]{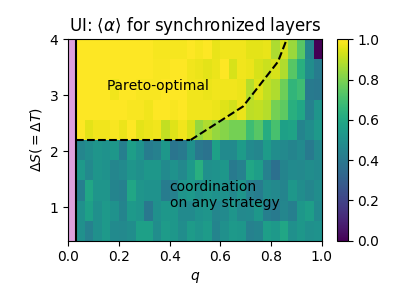}
    }
    \vspace{-0.5cm}
\caption{\footnotesize \textbf{UI in the diagonal case}
(i) Coordination rates on layers $\alpha^{I}$, $\alpha^{II}$,
and $\Delta \alpha$ vs node overlap $q$ for all
values of $\Delta S$.
(ii) Coordination rate $\alpha=\alpha^I=\alpha^{II}$
vs gap size $\Delta S$ for full node overlap $q=1$.
(iii) Phase diagram of coordination rate $\alpha=\alpha^I=\alpha^{II}$
in the $q$-$\Delta S$ space for synchronised layers.
Each layer has $N=1000$ nodes and forms a complete graph.
Averaged over (at least) 100 realisations.
}\end{figure}

\newpage
\begin{figure}[th!]
\centerline{
    \includegraphics[scale=0.6]{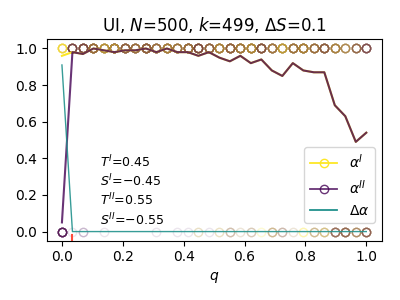}
    \includegraphics[scale=0.6]{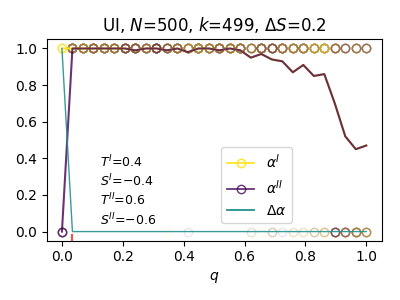}
    \includegraphics[scale=0.6]{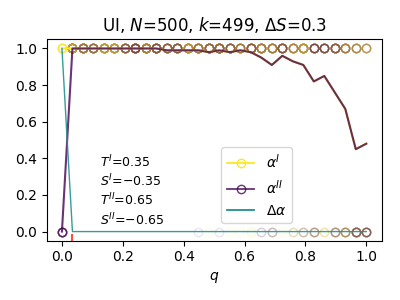}
    }
\centerline{
    \includegraphics[scale=0.6]{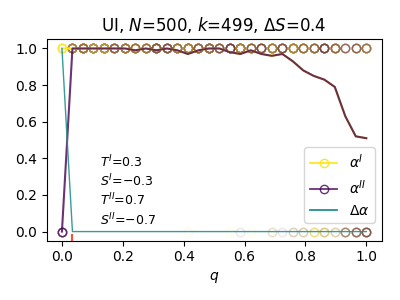}
    \includegraphics[scale=0.6]{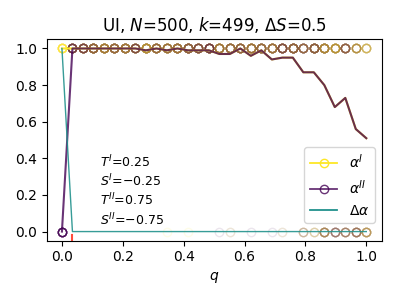}
    \includegraphics[scale=0.6]{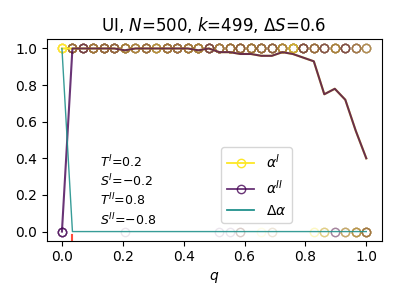}
    }
\centerline{
    \includegraphics[scale=0.6]{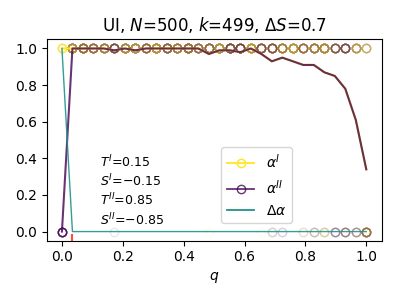}
    \includegraphics[scale=0.6]{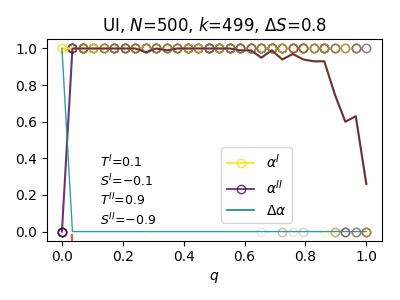}
    \includegraphics[scale=0.6]{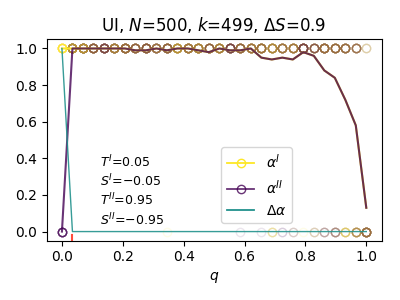}
    }
\centerline{
    \includegraphics[scale=0.6]{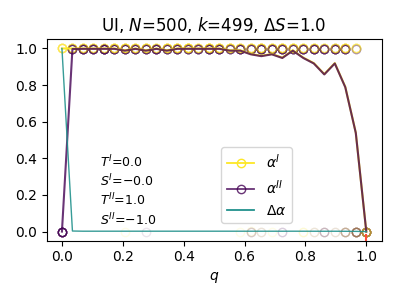}
    \includegraphics[scale=0.6]{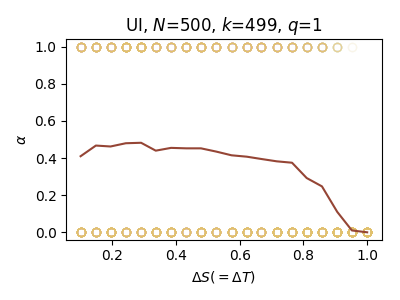}
    \includegraphics[scale=0.65]{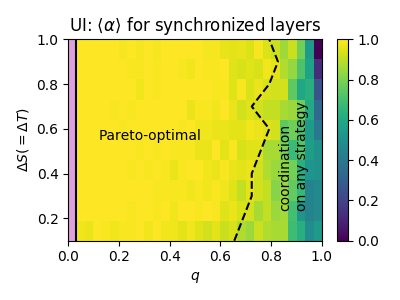}
    }
    \vspace{-0.5cm}
\caption{\footnotesize \textbf{UI in the stag hunt case}
(i) Coordination rates on layers $\alpha^{I}$, $\alpha^{II}$,
and $\Delta \alpha$ vs node overlap $q$ for all
values of $\Delta S$.
(ii) Coordination rate $\alpha=\alpha^I=\alpha^{II}$
vs gap size $\Delta S$ for full node overlap $q=1$.
(iii) Phase diagram of coordination rate $\alpha=\alpha^I=\alpha^{II}$
in the $q$-$\Delta S$ space for synchronised layers.
Each layer has $N=1000$ nodes and forms a complete graph.
Averaged over (at least) 100 realisations.
}\end{figure}

\newpage
\begin{figure}[th!]
\centerline{
    \includegraphics[scale=0.6]{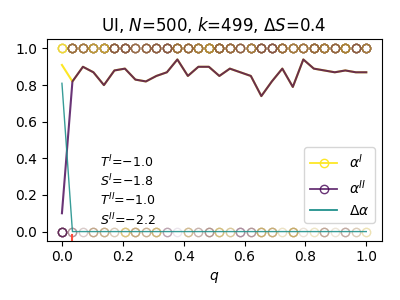}
    \includegraphics[scale=0.6]{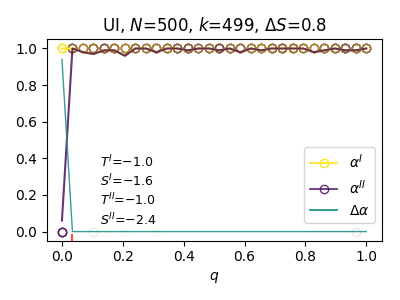}
    \includegraphics[scale=0.6]{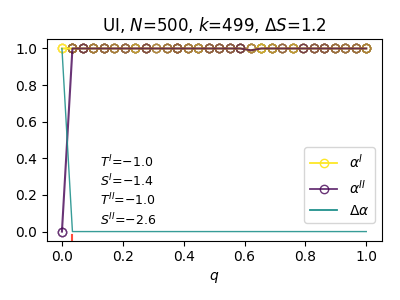}
    }
\centerline{
    \includegraphics[scale=0.6]{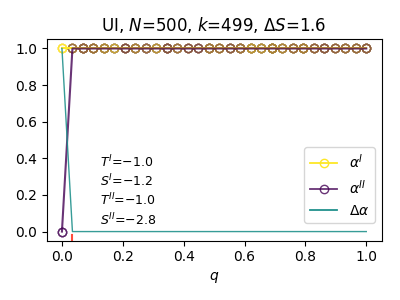}
    \includegraphics[scale=0.6]{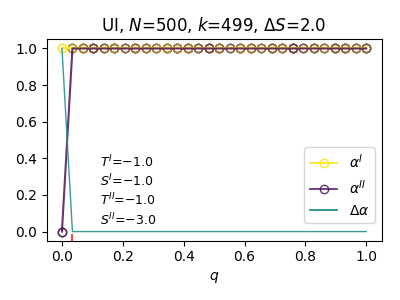}
    \includegraphics[scale=0.6]{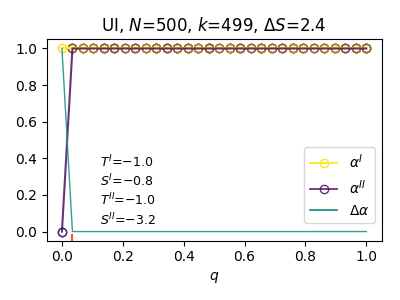}
    }
\centerline{
    \includegraphics[scale=0.6]{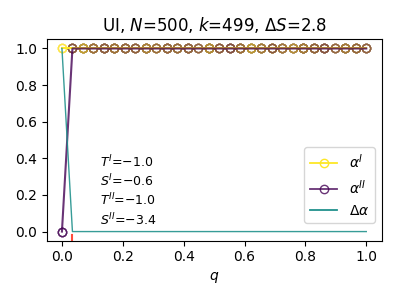}
    \includegraphics[scale=0.6]{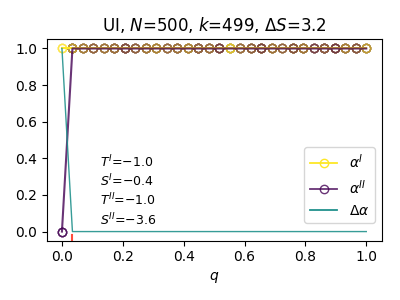}
    \includegraphics[scale=0.6]{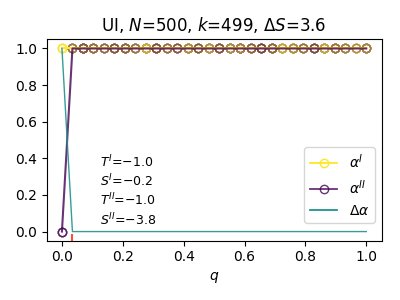}
    }
\centerline{
    \includegraphics[scale=0.6]{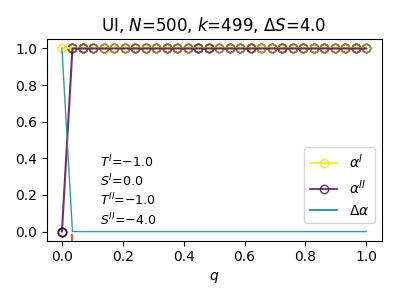}
    \includegraphics[scale=0.65]{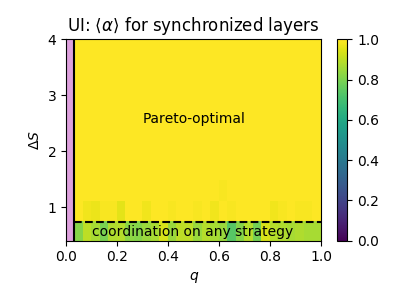}
    }
    \vspace{-0.5cm}
\caption{\footnotesize \textbf{UI in the horizontal case}
(i) Coordination rates on layers $\alpha^{I}$, $\alpha^{II}$,
and $\Delta \alpha$ vs node overlap $q$ for all
values of $\Delta S$.
(ii) Phase diagram of coordination rate $\alpha=\alpha^I=\alpha^{II}$
in the $q$-$\Delta S$ space for synchronised layers.
Each layer has $N=1000$ nodes and forms a complete graph.
Averaged over (at least) 100 realisations.
}\end{figure}

\newpage
\begin{figure}[th!]
\centerline{
    \includegraphics[scale=0.6]{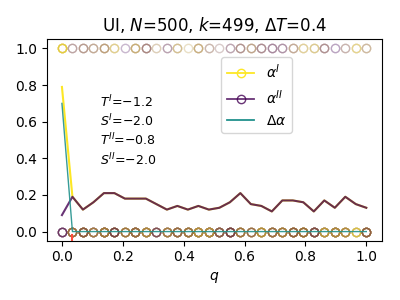}
    \includegraphics[scale=0.6]{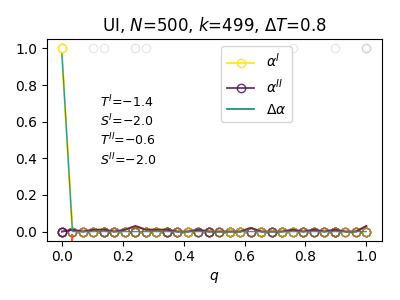}
    \includegraphics[scale=0.6]{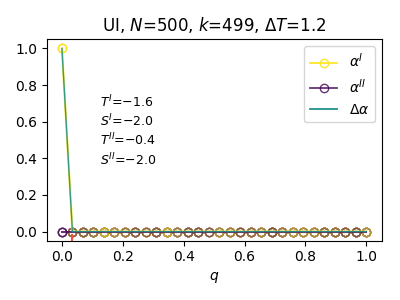}
    }
\centerline{
    \includegraphics[scale=0.6]{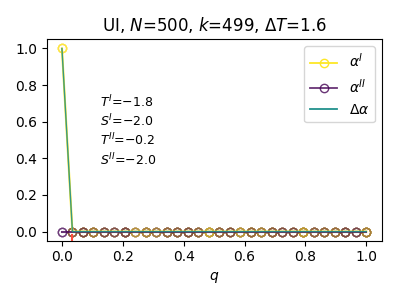}
    \includegraphics[scale=0.6]{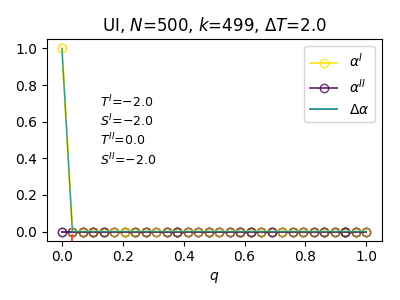}
    \includegraphics[scale=0.6]{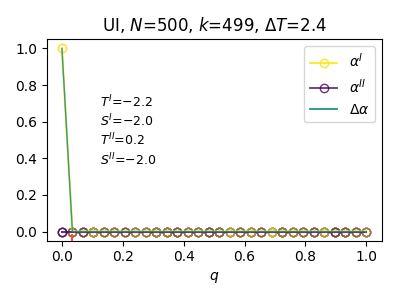}
    }
\centerline{
    \includegraphics[scale=0.6]{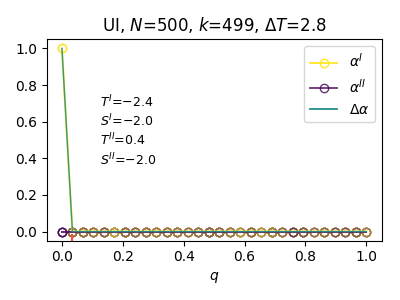}
    \includegraphics[scale=0.6]{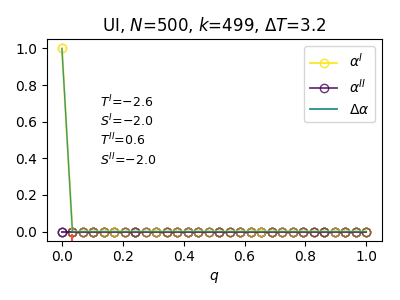}
    \includegraphics[scale=0.6]{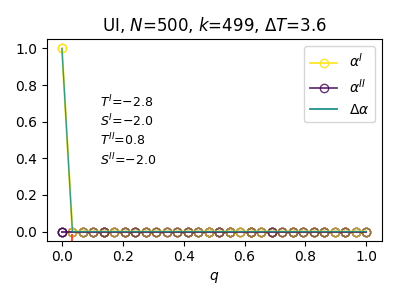}
    }
\centerline{
    \includegraphics[scale=0.6]{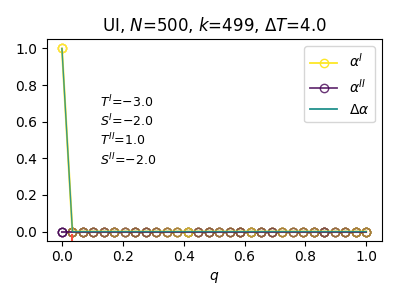}
    \includegraphics[scale=0.65]{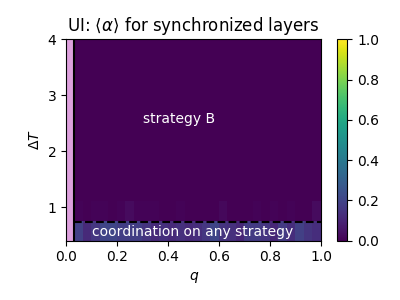}
    }
    \vspace{-0.5cm}
\caption{\footnotesize \textbf{UI in the vertical case}
(i) Coordination rates on layers $\alpha^{I}$, $\alpha^{II}$,
and $\Delta \alpha$ vs node overlap $q$ for all
values of $\Delta T$.
(ii) Phase diagram of coordination rate $\alpha=\alpha^I=\alpha^{II}$
in the $q$-$\Delta T$ space for synchronised layers.
Each layer has $N=1000$ nodes and forms a complete graph.
Averaged over (at least) 100 realisations.
}\end{figure}